\documentclass[12pt,preprint]{aastex}

\slugcomment{To appear in ApJ., 2006}

\shorttitle{MICROLENSING OF A BICONICAL BROAD LINE REGION}

\begin{document}

\title{MICROLENSING OF A BICONICAL BROAD LINE REGION}

\author{C. Abajas\altaffilmark{1},
E. Mediavilla\altaffilmark{1},  J. A. Mu\~noz\altaffilmark{2}, P. G\'omez-\'Alvarez\altaffilmark{1} and R. Gil-Merino\altaffilmark{3}}

\altaffiltext{1}{Instituto de Astrof\'\i sica de Canarias, V\'\i a L\'actea s/n,
E-38205 La Laguna, Tenerife, Spain; abajas@iac.es, pgomez@iac.es, emg@iac.es}
\altaffiltext{2}{Departament d'Astronomia i Astrof\'\i sica, Universidad  de
Valencia, Dr. Moliner 50, E-46100 Burjassot, Spain; jmunoz@uv.es}
\altaffiltext{3}{Institute of Astronomy, School of Physics, University of Sydney, NSW 2006, Australia; rodrigo@physics.usyd.edu.au}

\begin{abstract}

The influence of microlensing in the profiles of the emission lines generated in a biconical geometry is discussed. Microlensing amplification in this anisotropic model is not directly related to the bicone's intrinsic size but depends on the orientation of the bicone axis and on the cone aperture. The orientation of the projected bicone with respect to the shear of the magnification pattern can induce very interesting effects, like the quasi-periodic enhancements of the red/blue part of the emission line profile or the lack of correlation between the broad line region (BLR) and continuum light curves of QSOs. The emission line profiles of a BLR moving in a high caustic concentration exhibit sharp features that are well defined in wavelength. These features (spikes) correspond to the scanning of the kinematics of the BLR by the caustic clusters. The biconical model can qualitatively reproduce with a transversal (with respect to the shear) movement of the BLR, the recurrent blue-wing enhancement detected in the emission line profile of the A image of the quasar lensed system SDSS J1004+4112. The probability of observing this repetitive event is almost a 2\% for a fraction of matter in stars of a 5\%. This result would make plausible the detection of the spectral variability in SDSS J1004+4112 under the hypothesis of microlensing of a bicone.

\end{abstract}

\keywords{cosmology: gravitational lensing --- quasars: emission lines}

\section{INTRODUCTION}

The variability of the light-curve of a lensed object induced by the discretization of the mass distribution of the lens galaxy in stars or other compact objects (gravitational microlensing, Chang \& Refsdal 1979; Young 1981; Paczynski 1986; Kayser, Refsdal \& Stabell 1986; Schneider \& Weiss 1987; Irwin et al. 1989; Witt 1990; Wambsganss 2001; Gil-Merino \& Lewis 2005) offers a promising way of studying the unresolved source of the continuum of QSOs. Other characteristic regions of QSOs and AGN, such as the broad line region (BLR) and the narrow line region (NLR), were considered too large (Rees 1984) to be significantly affected by the microlensing effect (Nemiroff 1988; Schneider \& Wambsganss 1990). However, recent studies by Abajas et al. (2002) and Lewis \& Ibata (2004) have shown that, according to the new estimates by Wandel, Peterson \& Malkan (1999), the size of the BLR is small enough to be significantly amplified by microlensing of stellar-sized objects. Other AGN components such as the broad absorption line region may also be studied by microlensing (e.g. Belle \& Lewis 2000; Chelouche 2005).

In studies by Abajas et al. (2002) and Lewis \& Ibata (2004) a number of plausible  geometries for the BLR, taken from the complete analysis of Robinson (1995), have been considered. These models include, in decreasing order of symmetry, spherical, cylindrical and biconical geometries, which are combined with different emissivity and velocity laws. Additional models for the BLR could be found in Murray et al. 1995.~An expected result confirmed by Abajas et al. (2002) and Lewis \& Ibata (2004) is that the less symmetrical models induce the most interesting distortions in the line profiles. However, the microlensing of the biconical model has been studied in only two limiting cases, for inclinations 0$^\circ$ and 90$^\circ$, in which the crucial effects of differential amplification between the receding and approaching parts of the velocity field are absent. Another important characteristic of the biconical model is that the length of the cones is not a direct estimate of the size of its projected surface and, hence, that the microlensing (de-)amplification can be strongly dependent on the orientation and aperture of the cones.

In this paper we explore microlensing effects on a biconical model consistent with much experimental evidence (Marziani,
Calvani, \& Sulentic 1992; Wilson \& Tsvetanov 1994) and theoretical work supporting the idea that the flow of emitting gas in AGN is anisotropic, preferentially confined to a pair of oppositely directed cones (Zheng, Binette, \& Sulentic 1990; Robinson 1995). We consider the microlensing effects at high optical depth on the profiles of the broad emission lines (BEL). In \S3 the effects on the continuum are compared with those on the BLR. We then analyze the possible use of this model in explaining the characteristic variability detected in the BEL of the quadruple quasar lensed system SDSS J1004+4112.

\section{MICROLENSING EFFECTS AT HIGH OPTICAL DEPTH ON THE PROFILE OF BROAD
EMISSION LINES}

Throughout this paper we use the standard notation for gravitational
lensing (Schneider, Ehlers, \& Falco 1992).
All microlenses have a common mass, $1M_{\odot}$, and
all distances are normalized to the Einstein radius associated with this
microlens mass. The Einstein radius, $\eta_0$, projected onto the source plane is:
\begin{equation}
\eta_0=\sqrt{\frac{4GM}{c^2}\frac{D_{\rm s}D_{\rm ds}}{D_{\rm d}}},
\end{equation}
where $G$ is the gravitational constant, $M$ is the single mass of the stars,
$c$ is the speed of light, and
$D_{\rm d}$, $D_{\rm s}$ and $D_{\rm ds}$ are the angular diameter distances to
the lens, to the source, and from the lens to the source respectively. The Einstein radius projected onto the lens plane is $\xi_0=\eta_0 D_{\rm d}/D_{\rm s}$.

 A standard cosmology is assumed with 
$H_0=70~{\rm kms}^{-1} {\rm Mpc}^{-1}$, $\Omega_{\rm M}=0.3$, and 
$\Omega_{\Lambda}=0.7$.

\subsection{Magnification patterns}

We have selected as an illustrative example for a general study of the microlensing of a biconical BLR the gravitational lens  discovered by Jackson
et al. (1995), B 1600+434, a double-imaged quasar  at $z_{\rm s}=1.59$,
lensed by an edge-on spiral galaxy at $z_{\rm l}=0.42$. This is a well studied
lens system that Abajas et al. (2002) found to be a good candidate for observing
microlensing of the BLR. Given the low number of constrains for this double-imaged quasar we just fitted a singular isothermal sphere plus an external shear to estimate the microlensing parameters, the convergence or optical depth, $\kappa$, and the shear due to the external mass, $\gamma$. For A and B images the values for
$(\kappa,\gamma)$ are (0.30,0.16) and (0.80,0.72) respectively.

The normalized lens equation for a field of point masses with an external shear, $\gamma$, and a smooth mass distribution, $\kappa_{\rm c}$,
where the coordinate frame has been oriented such that the shear acts
along one of the coordinate axes, is
\begin{equation}
{\bf y}=\left ( \begin{array}{cc}
1-\kappa_{\rm c}+\gamma & 0 \\
0        & 1-\kappa_{\rm c}-\gamma \\
\end{array}\right ){\bf x} + \sum_{i=1}^{N_*} m_i
\frac{({\bf x}-{\bf x}_i)}{|{\bf x}-{\bf x}_i|^2},
\end{equation}
where ${\bf x}$ and ${\bf y}$ are the normalized image and source positions,
respectively, and ${\bf x}_i$ and $m_i$ are the normalized positions and
masses of microlenses. The total surface mass density or optical depth can be written as $\kappa=\kappa_* + \kappa_{\rm c}$, where $\kappa_*$ represents the contribution from the compact microlenses.

\begin{figure}
\figurenum{1}
\plotone{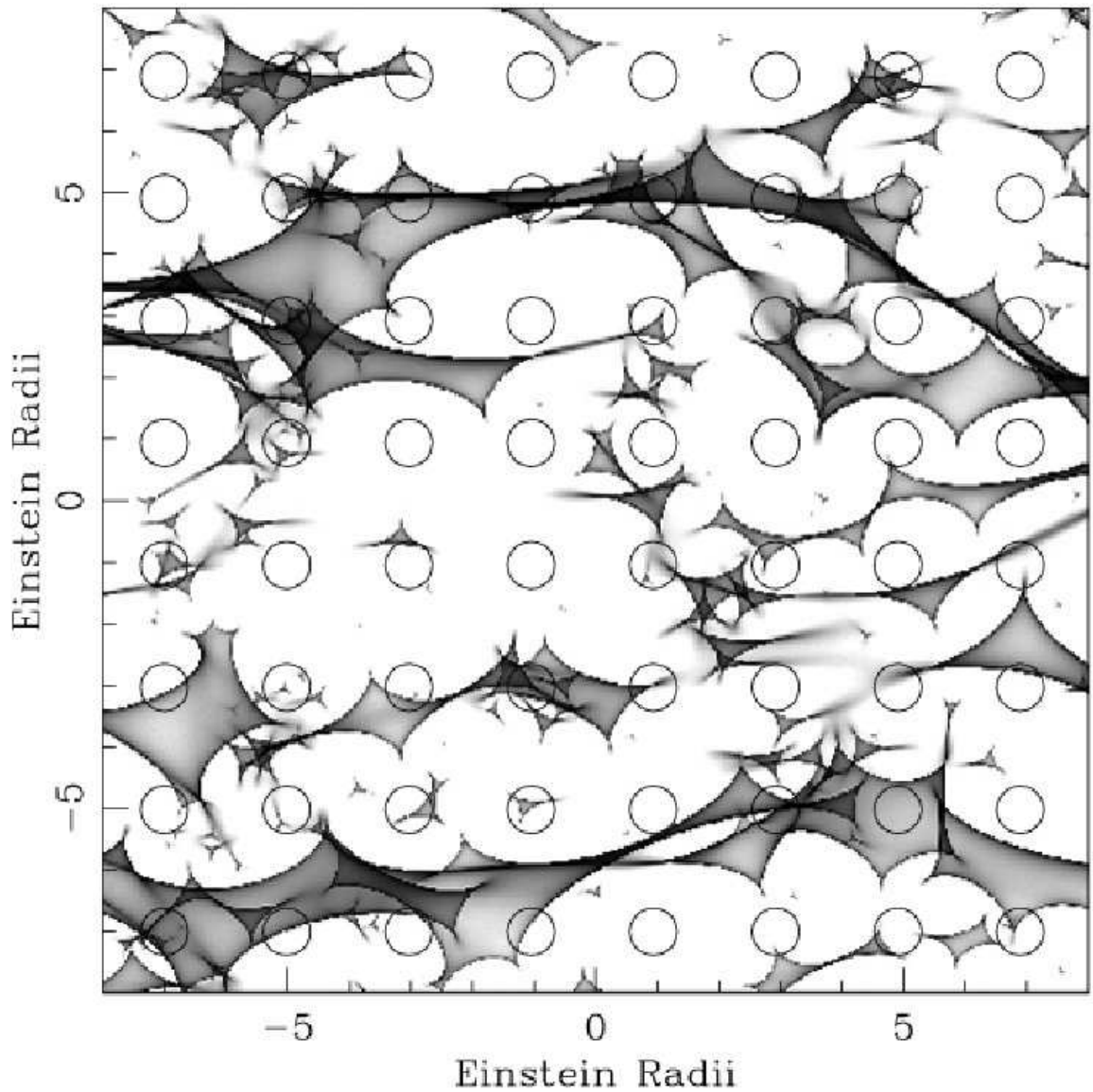}
\figcaption[]{The magnification map, $\mu (\vec{r})$, for the B image of B1600+434, used here to calculate the changes in the shape of the line profile. Each circle represents the BLR in different positions over the magnification map with $r_{\rm ex}=0.37 \eta_0$.\label{fig1}}
\end{figure}

A ray-shooting method was used to compute the magnifications maps (Kayser, Refsdal \& Stabell 1986; Wambsganss 1990).
The map of the B1600+434B image (Fig.~\ref{fig1}), with
an area of $16\eta_0\times 16\eta_0$ at the source plane, was computed covering
an area of $60\xi_0\times 60\xi_0$ in the lens plane, and
was obtained at a resolution of $1000\times1000$ pixels.
256 rays were shot per unlensed pixel, which means $3.6\cdot10^9$ rays
traced between the observer and the source.
The number of stars considered was 912, assuming that the entire optical depth is in compact objects, $\kappa_{\rm c}=0$.
The theoretical mean amplification $<\mu_{th}>=|(1-\kappa)^2-\gamma^2|^{-1}$ is
2.068, and the mean amplification of the map is $<\mu>=1.900$.

\subsection{Biconical Broad Emission Line Region}

In this section we adopt the kinematic model for a biconical BLR used by
Robinson (1995) to study the effects of microlensing over the line profile shapes in the
context of a simple parameterization of some basic properties of the BLR, such 
as emissivity, velocity law, the cone aperture and inclination. 

We compute the broad emission-line profile, which depends on the wavelength $\lambda$, from the expression
\begin{equation}
\label{fluxl}
F _\lambda=\int_V \epsilon(r) ~ \delta \left [\lambda-\lambda_0
\left (1+\frac{v_{\shortparallel}}{c}\right )\right ] ~ \mu(\vec{r}) ~ dV ,
\end{equation}
where $\mu$ is the amplification over the caustics pattern, $\delta$ is the delta function, $V$ is the total volume, $c$ is the speed of light, and the emissivity
and magnitude of the velocity follow radial dependences as:
\begin{equation}
\label{ecemis}
\epsilon(r)=\epsilon_0 \left (\frac{r}{r_{\rm in}}\right )^{\beta}
\end{equation}
and
\begin{equation}
\label{velocity}
v(r)=v_0\left (\frac{r}{r_{\rm in}}\right )^{p}.
\end{equation}

We adopt inner ($r_{\rm in}$) and outer ($r_{\rm ex}$) radii for the BLR.

\begin{figure}
\figurenum{2}
\plottwo{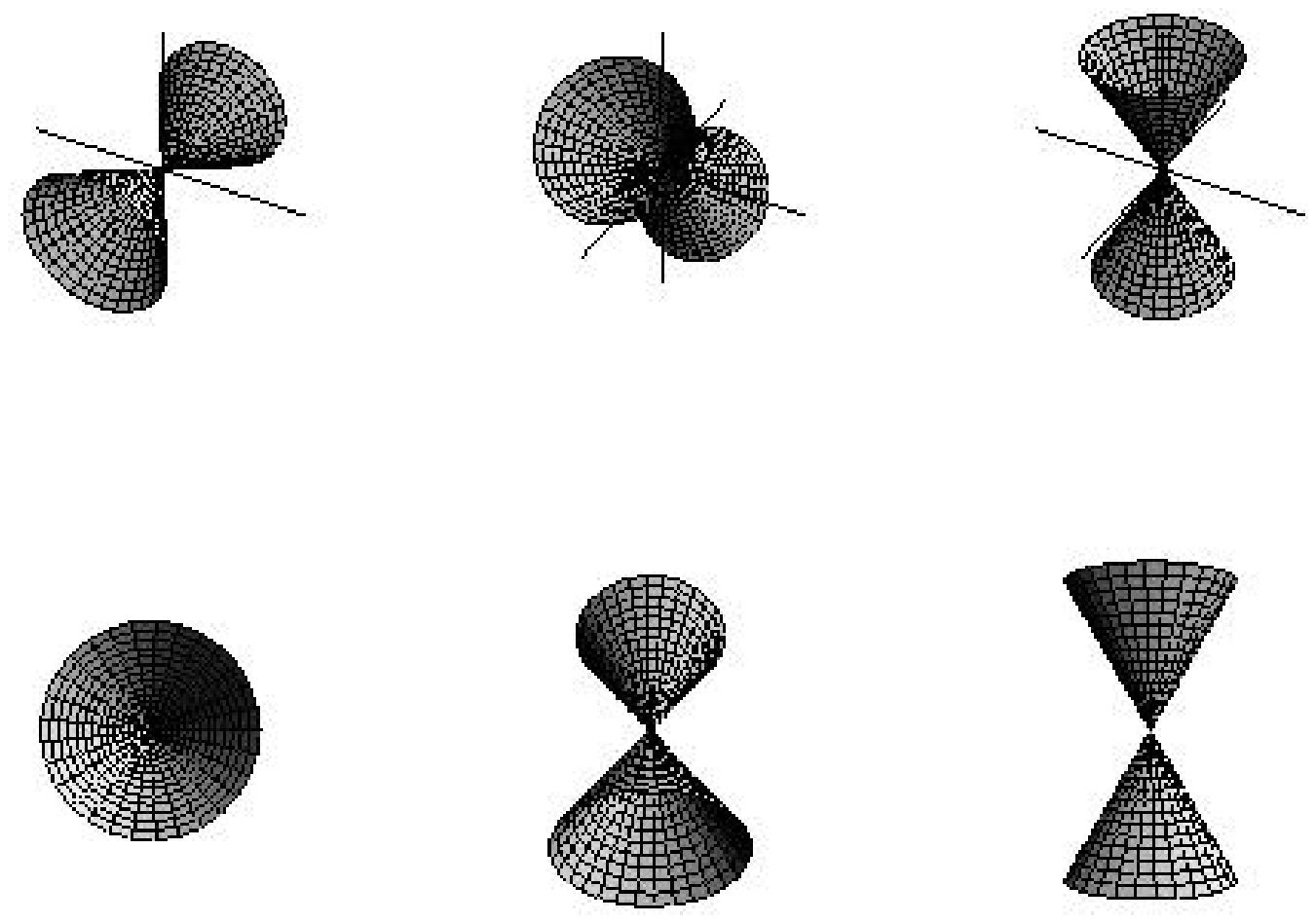}{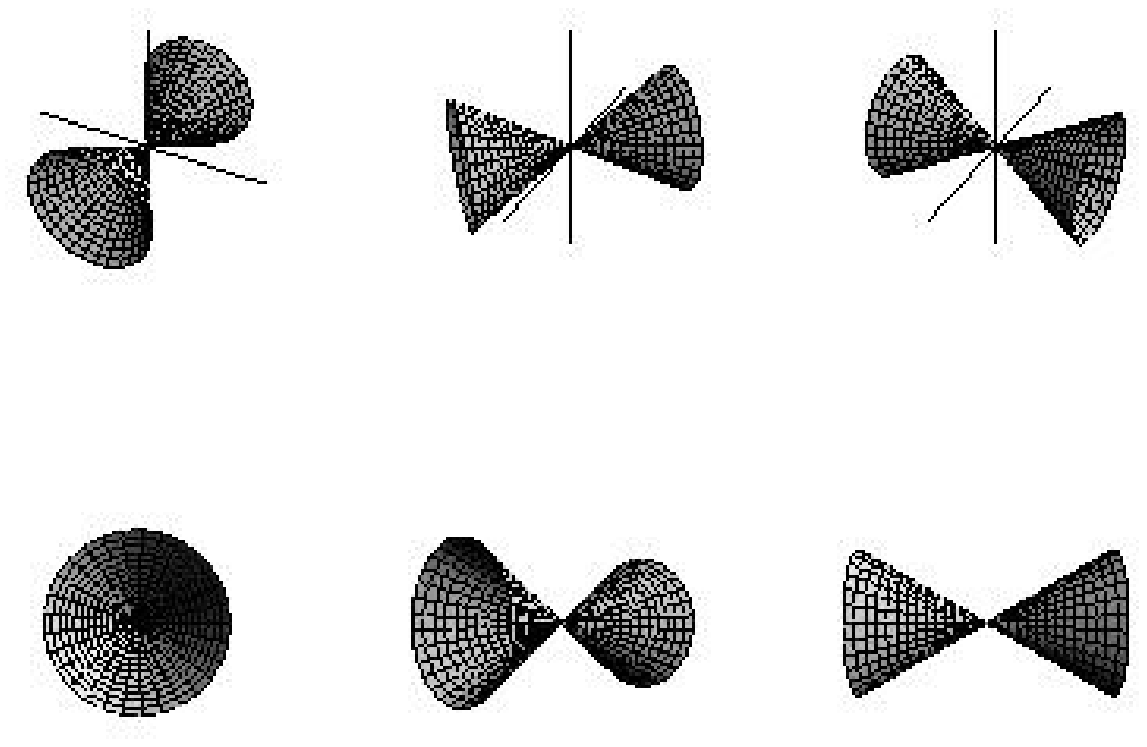}
\figcaption[]{Projections of the bicone, perpendicular and parallel, from left to right with a cone half-aperture $\theta_c=30^\circ$.\label{fig2}}
\end{figure}

In a biconical model we need three polar coordinates $(r,\theta,\varphi)$ 
measured with respect to the cone axis to express the projected
line-of-sight velocity corresponding to an emitter,
\begin{equation}
v_{\shortparallel}=v_0 \left (\frac{r}{r_{\rm in}}\right )^p
\xi_{\perp,\parallel}\quad (p>0),
\end{equation}
with
\begin{mathletters}
\begin{equation}
\xi_{\perp}=\sin \theta \sin \varphi \sin i + \cos \theta \cos i,
\end{equation}
\begin{equation}
\xi_{\parallel}=\sin \theta \cos \varphi \sin i + \cos \theta \cos i,
\end{equation}
\end{mathletters}
where $i$ is the inclination of the cones axis with respect to the line of
sight, and $\xi_{\perp}$ and $\xi_{\parallel}$ are the projections of the bicone when it turns around the x and y axes, respectively. Both projections fit when $i=0^\circ$. Figure~\ref{fig2} shows the perpendicular projection on the top and the parallel projection on the bottom.

Following the Appendix \ref{apen}, the line profile is obtained by integrating
\begin{equation}
 F_x  = \left\{
  \begin{array}{ll}
	a\left (\frac{x}{x_{\rm m}}\right)^{\eta}
	\int\limits_{0}^{2\pi}\left (\left [
        \int\limits_{0}^{\theta_{\rm c}} +
        \int\limits_{\pi-\theta_{\rm c}}^{\pi}\right ]                  
        \frac{[\mu(\vec{r})]_{\perp,\parallel}^{\rm f=0}}
	{\left| \xi_{\perp,\parallel} 
	\right |^{\eta +1}} ~\sin\theta d\theta \right )d\varphi,
  & \mbox {\small for $x_{\rm m}|\xi_{\perp,\parallel}|<|x|<|\xi_
  {\perp,\parallel}|$} \\
0 & \mbox {in other cases.}  
  \end{array}
  \right.        
\end{equation}
where $x=c(\lambda-\lambda_0)/(v_{\rm max}\lambda_0)$, $a=(\epsilon_0 r_{\rm in}^3 c)/(\lambda_0 p v_0)$, $x_{\rm m}=v_0/v_{\rm max}=(r_{\rm in}/r_{\rm ex})^p$, $\theta_{\rm c}$ is the cone aperture half-angle, $\eta= (\beta+3-p)/p$ and
\begin{mathletters}
\begin{equation}
[\mu(\vec{r})]_{\perp}^{\rm f=0}= \mu \left ([r]_{\perp}^{\rm f=0}  \sin \theta 
\cos \varphi ,[r]_{\perp}^{\rm f=0} (\sin \theta \sin \varphi \cos i - \cos
\theta \sin i) \right )
\end{equation}
\begin{equation}
[\mu(\vec{r})]_{\parallel}^{\rm f=0}= \mu \left ([r]_{\parallel}^{\rm f=0}
( \sin \theta \cos \varphi \cos i - \cos \theta \sin i ),
[r]_{\parallel}^{\rm f=0} \sin \theta \sin \varphi   \right).
\end{equation}
\end{mathletters}

\subsection{Estimate of the BLR Radius}

To estimate the BLR radius for B1600+434 we apply the relationship $r_{\rm BLR}\propto L^{0.67}$ from Kaspi et al. (2005) using NGC 5548 as a reference object. We have adopted the BLR size for high ionization lines of NGC 5548 as 2.5
light-days (He~{\sc ii} line, Peterson \& Wandel 1999) and from NED the visible
magnitude and redshift of NGC 5548, $m_V=13.3$ and $z=0.017$. To
transform apparent magnitudes into intrinsic luminosities we have made use
of the equation $L=4\pi D_{\rm lum}^{2} S (1+z)^{(\alpha-1)}$, which relates
the absolute luminosity of the source, the luminosity distance,
$D_{\rm lum}$, the apparent flux, $S$, and the spectral index, $\alpha$,
defined by $F_{\nu}\propto \nu^{-\alpha}$. We have considered a spectral index
$\alpha=0.5$ (e.g., Richards et al. 2001). This is the procedure followed by Abajas et al. (2002) where, from an estimated intrinsic magnitude for B1600+434 of $m_V=22.69$, the BLR radius for this double-imaged quasar is $r_{\rm BLR}=4.95$ l-days for high ionization lines.

This is a reverberation mapping size obtained, through Kaspi et al. (2005) relationship, from observations of many objects. In principle, we could suppose that the BLRs of these objects are bicones with apertures ranging from 0 to 180 degrees (spherical case) and with random orientations. If the bicone axis is pointing towards the observer the reverberation size will depend on the bicone aperture and will approximate zero for small apertures. If the bicone axis were perpendicular to the observer line of sight, the reverberation sizes will be luminosity-weighted radii (Koratkar \& Gaskell 1991) that correspond (Robinson \& P\'erez 1990) in the case of a biconical geometry to
\begin{equation}
r_{\rm BLR}=\frac{\int_V r\epsilon (r) dV}{\int_V \epsilon (r) dV}=
\frac{\int_{r_{\rm in}}^{r_{\rm ex}} r^{\beta+1}4\pi (1-\cos \theta_{\rm c}) r^2 dr}{\int_{r_{\rm in}}^{r_{\rm ex}}r^{\beta} 4\pi (1-\cos \theta_{\rm c}) r^2 dr}=
\frac{\beta+3}{\beta+4}~~ ~\frac{1-(r_{\rm in}/r_{\rm ex})^{\beta +4}}{1- (r_{\rm in}/r_{\rm ex})^{\beta +3}}~ r_{\rm ex},
\end{equation}
that is, independent of the cone aperture and, hence, coincident with the spherical case. In principle we should average over all the possible inclinations and cone apertures of the objects considered to derive the relationship of Kaspi et al. (2005) to infer the outer geometrical radius of the BLR ($r_{\rm ex}$). However, having into account that the bicone orientation perpendicular to the observer is the most probable one and that we lack on information about the bicone aperture we finally adopt the perpendicular case (that includes the spherical case) with an emissivity law $\beta=-1.5$ and
$r_{\rm in}=0.1\ r_{\rm ex}$. Then $r_{\rm ex}=r_{\rm BLR}/0.618=0.37\eta_0$, with $1\eta_0 (1M_{\odot})=21.58$ l-days in this gravitational lens. This is, thus, an estimate of the intrinsic geometric size (probably slightly underestimated) of the BLR of one object of luminosity equal to that of B1600+434.  Depending on the orientation and aperture of the bicone, this intrinsic size will result in different projected areas in the observer line of sight and, hence, in different microlensing amplifications.

\subsection{Microlensed Line Profiles}

Using Eq.~8 we have computed microlensed line profiles at different positions on the magnification pattern (Fig.~\ref{fig1}). Different values for the inclination,
$i=0^\circ,45^\circ,90^\circ$ and the cone aperture half-angle, 
$\theta_{\rm c}=10^\circ,30^\circ$, are considered, as well as two different exponents for the velocity law, $p=0.5,2$.

\begin{figure}
\figurenum{3}
\plottwo{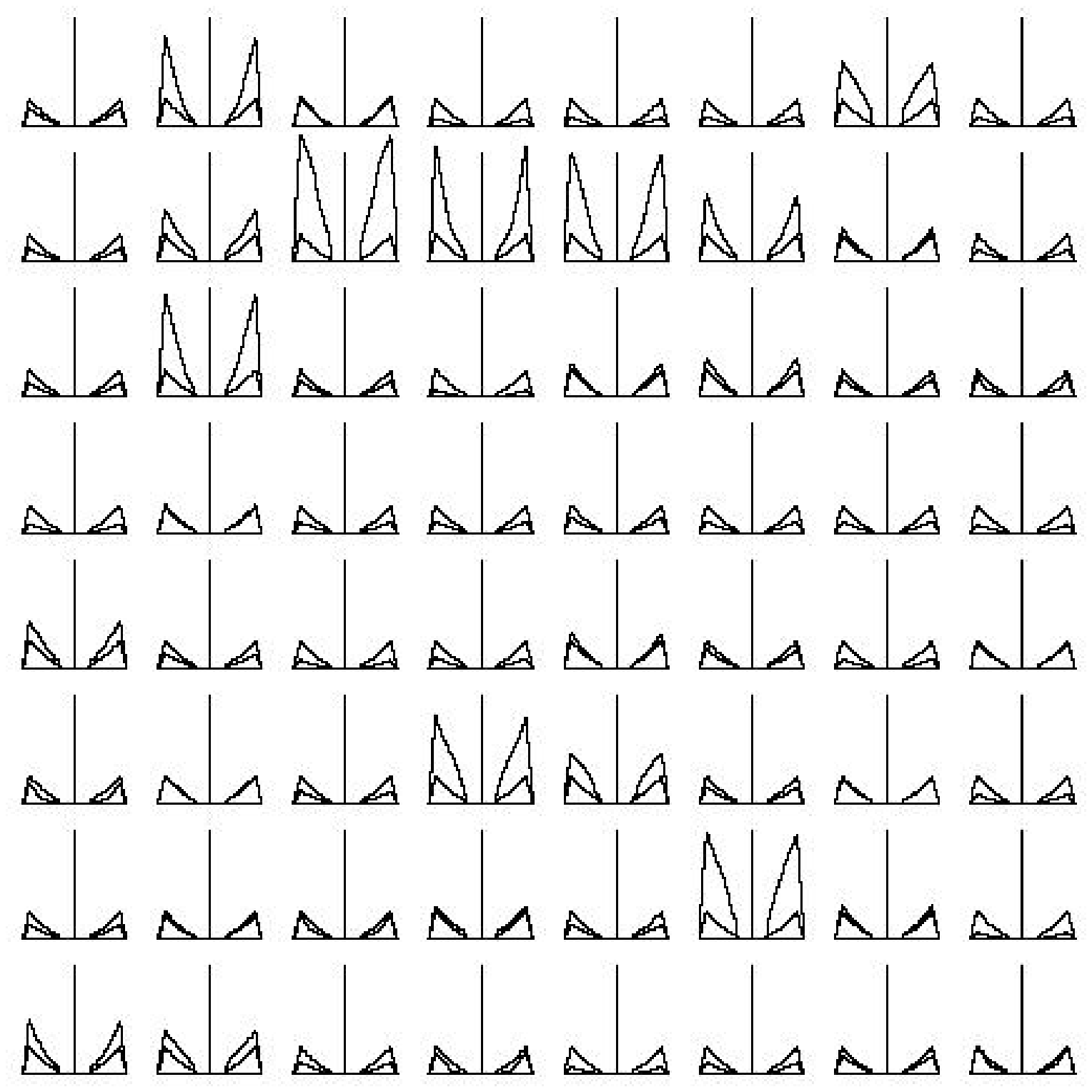}{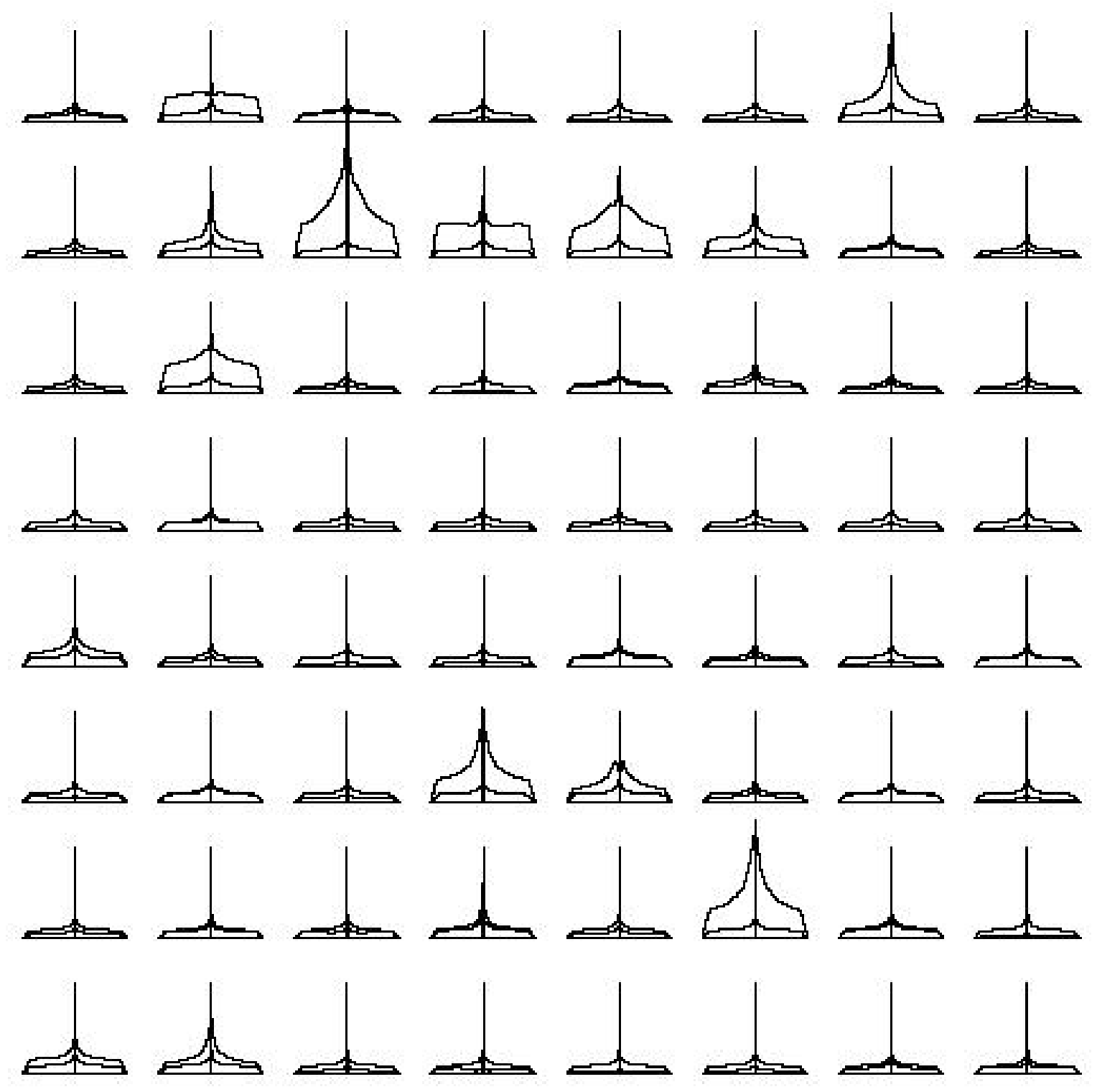}
\figcaption[]{Biconical model with $i=0^{\circ}$, $\theta_c=30^\circ$, $p=0.5$ for the left panel and $p=2$ for the right panel. The heavy solid line is the amplified line profile and the lighter solid line is the line profile normalized to mean amplification in the magnification map (Fig.~\ref{fig1}).\label{fig3}}
\end{figure}

Figure~\ref{fig3} corresponds to the $i=0^\circ$ case with a cone half-aperture of $\theta_{\rm c}=30^\circ$. The exponent of the velocity law, $p$, is 0.5 and 2.0 for the left and right panels, respectively. In this limiting case the bicone is projected along its axis and would appear projected as a circle of radius $r_{ex}\sin \theta_{\rm c}$. We obtain both strong amplification and significant deamplification of the line emission at different positions in the source plane.

\begin{figure}
\figurenum{4}
\plottwo{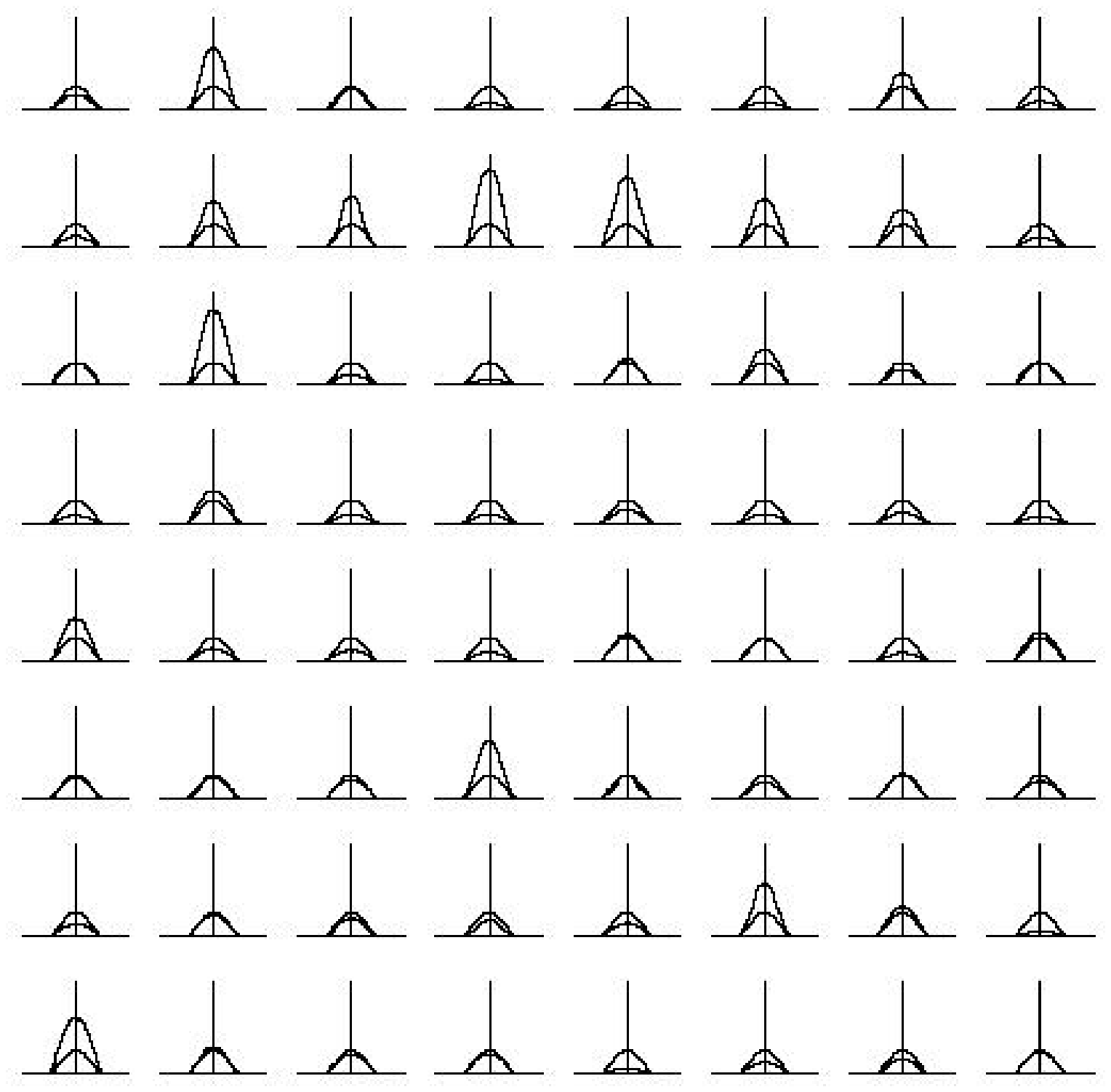}{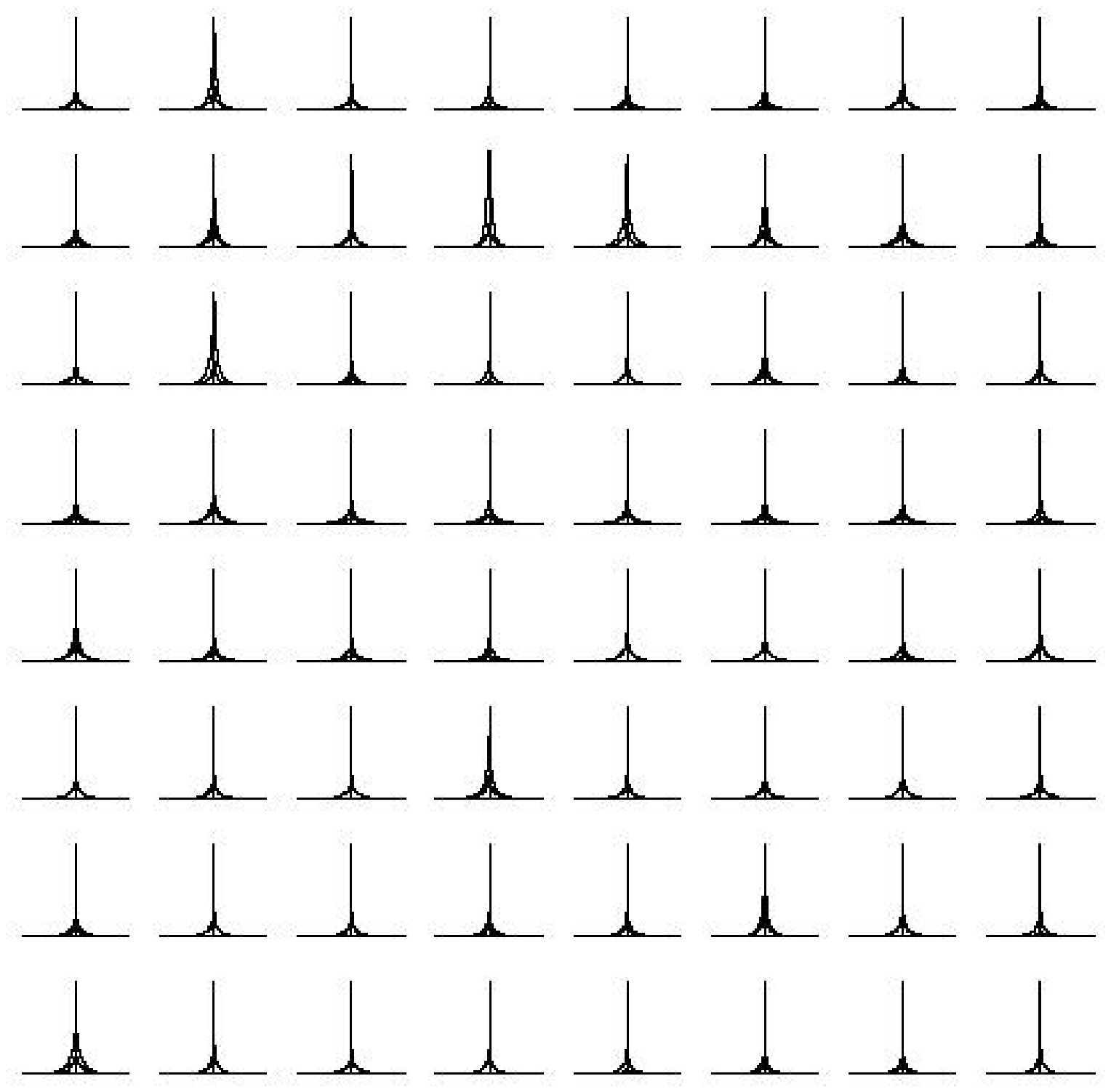}

\plottwo{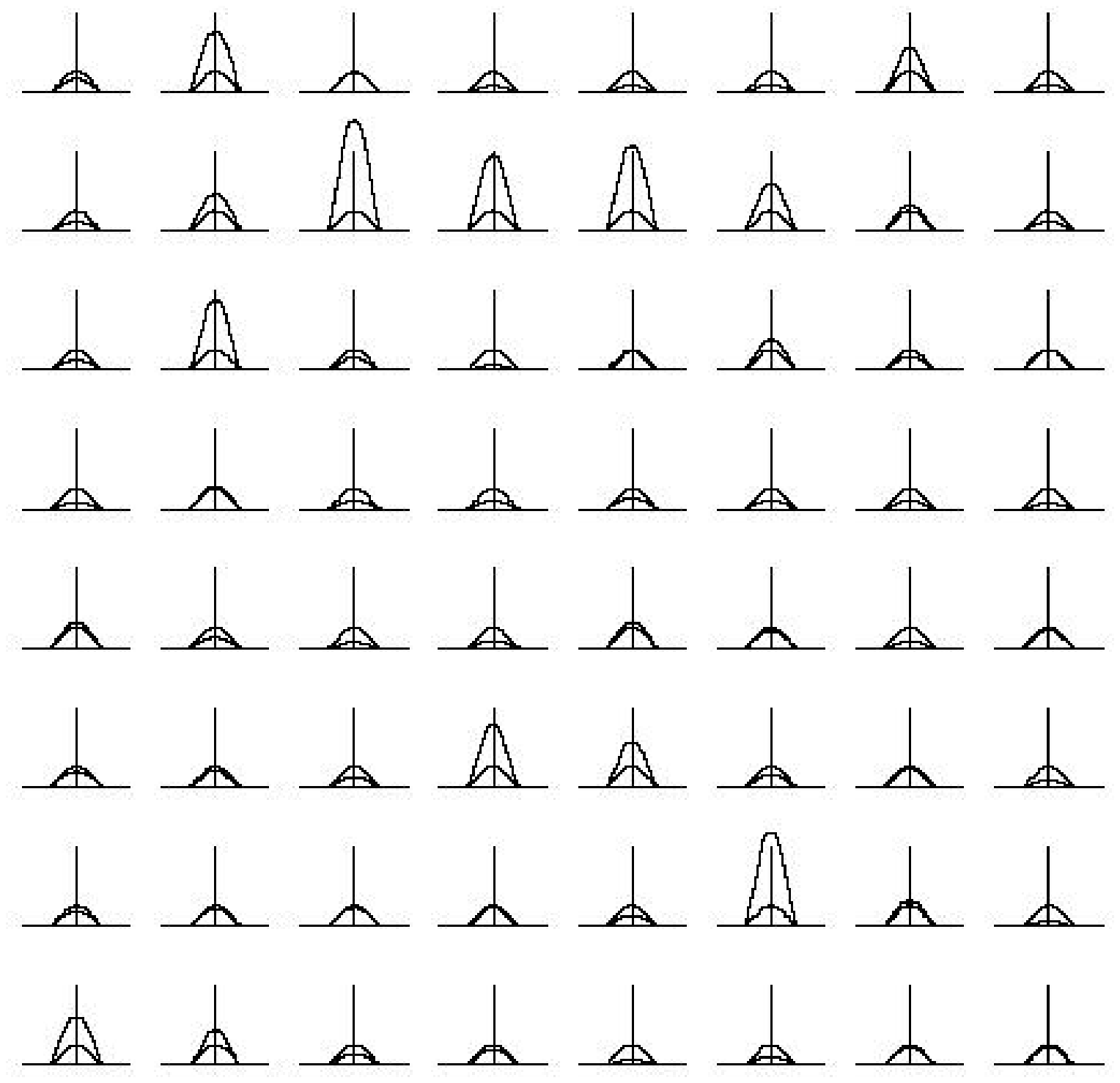}{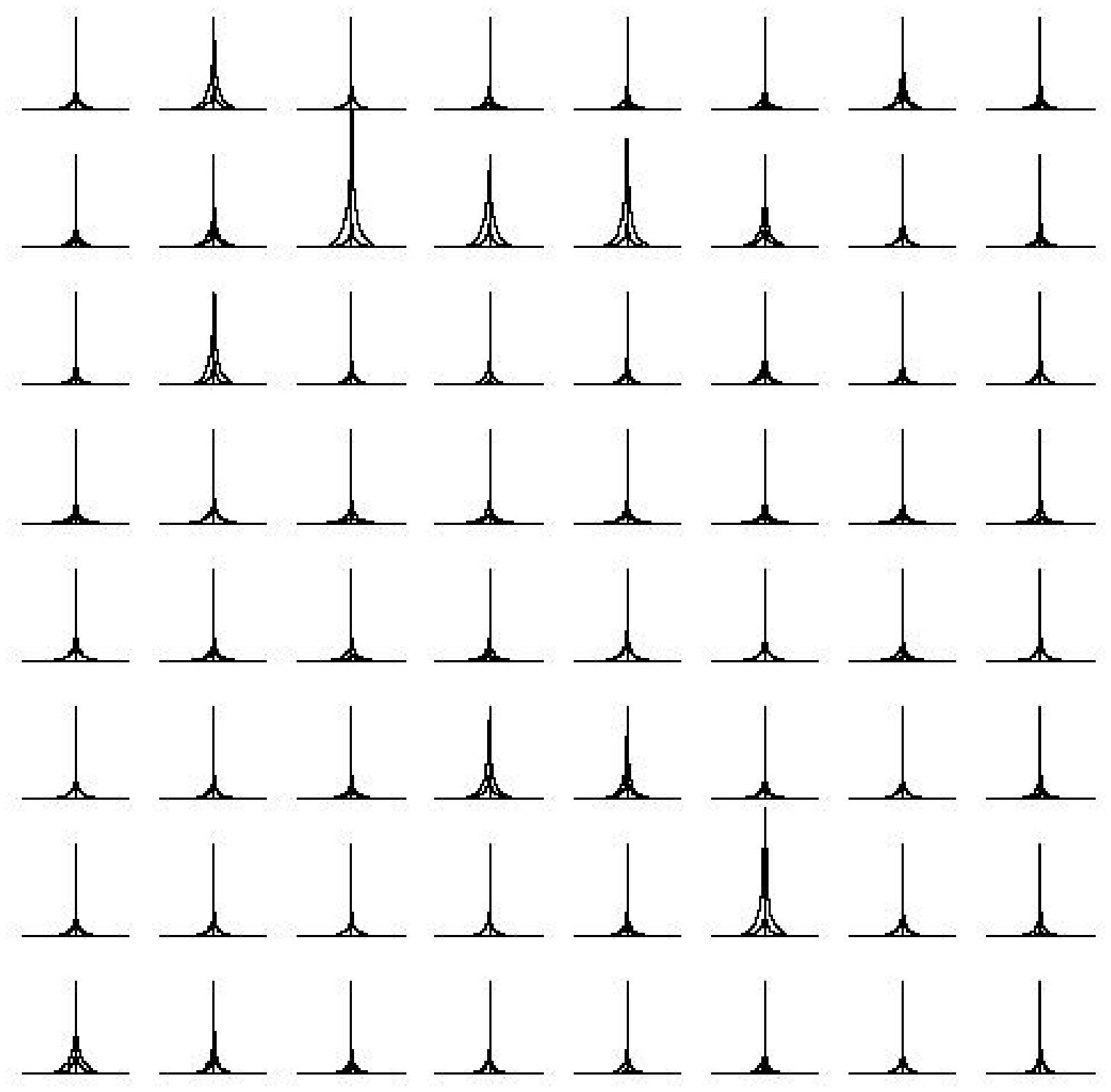}
\figcaption[]{Biconical model with $i=90^{\circ}$, $\theta_c=30^\circ$, $p=0.5$ for the left panels and $p=2$ for the right panels. The perpendicular projections of the velocity correspond to the upper panels and the parallel projections correspond to the lower panels. The heavy solid line is the amplified line profile and the lighter solid line is the line profile normalized to mean amplification in the magnification map (Fig.~\ref{fig1}).\label{fig4}}
\end{figure}

Figure~\ref{fig4} corresponds to the $i=90^\circ$ case with a cone half-aperture of $\theta_{\rm c}=30^\circ$ (notice that now the main component of the velocity is perpendicular to the line of sight and the line profile has only one peak). In this other limiting case the bicone is projected perpendicular to its axis and would appear as a "bi-triangle".  It is evident that for this inclination microlensing effects on the line profile depend on the orientation of the bicone axis with respect to the caustics when the source moves across the magnification pattern. In the absence of any privileged direction in the magnification pattern this would result in a random effect. However, in presence of shear, the orientation of the cone axis induces very interesting effects. In Fig.~\ref{fig4}, for instance, a greater number of strong amplification events are found when the cone axis is parallel to the shear (lower panels) than when it is perpendicular (upper panels).

\begin{figure}
\figurenum{5}
\plottwo{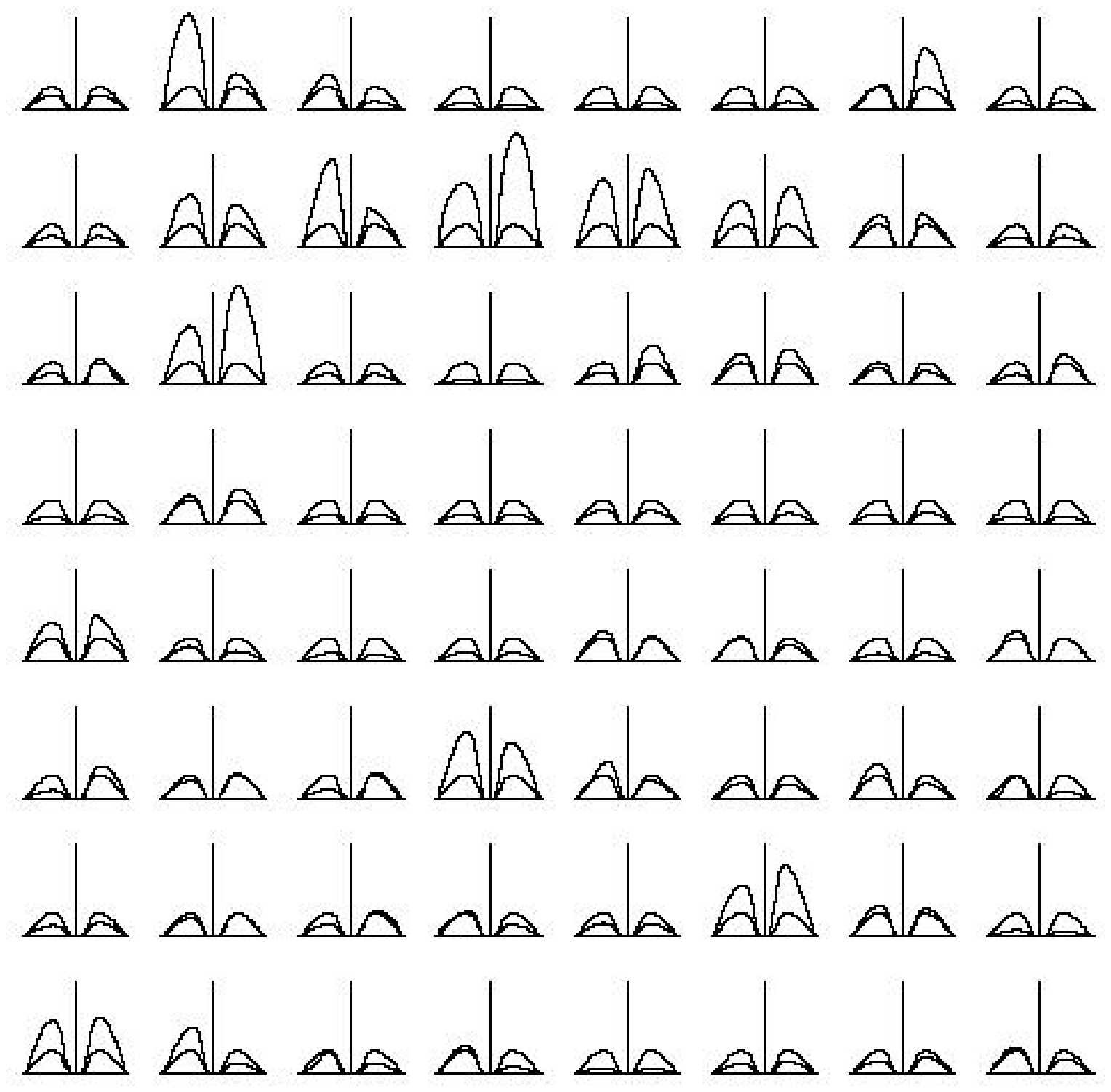}{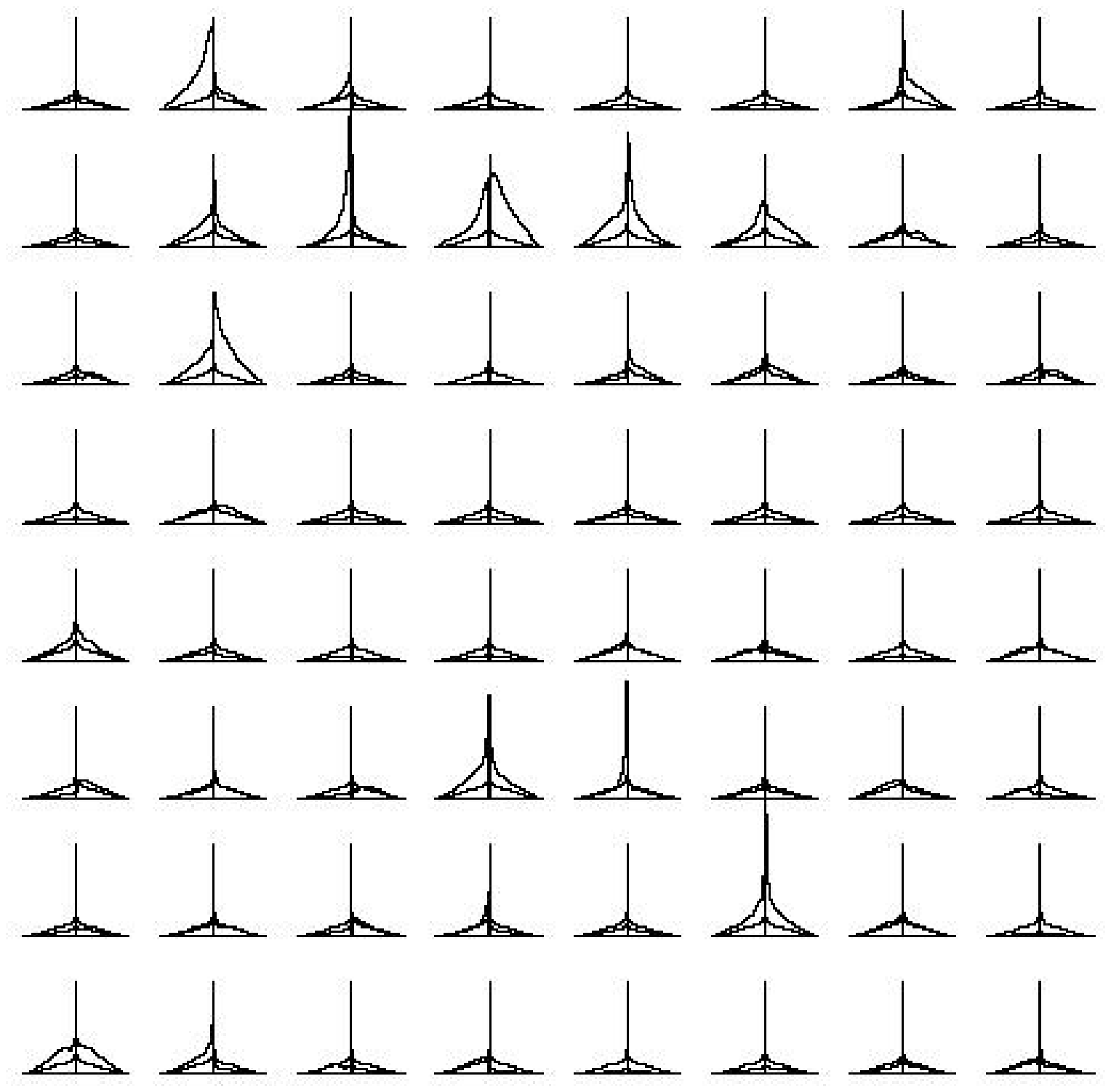}

\plottwo{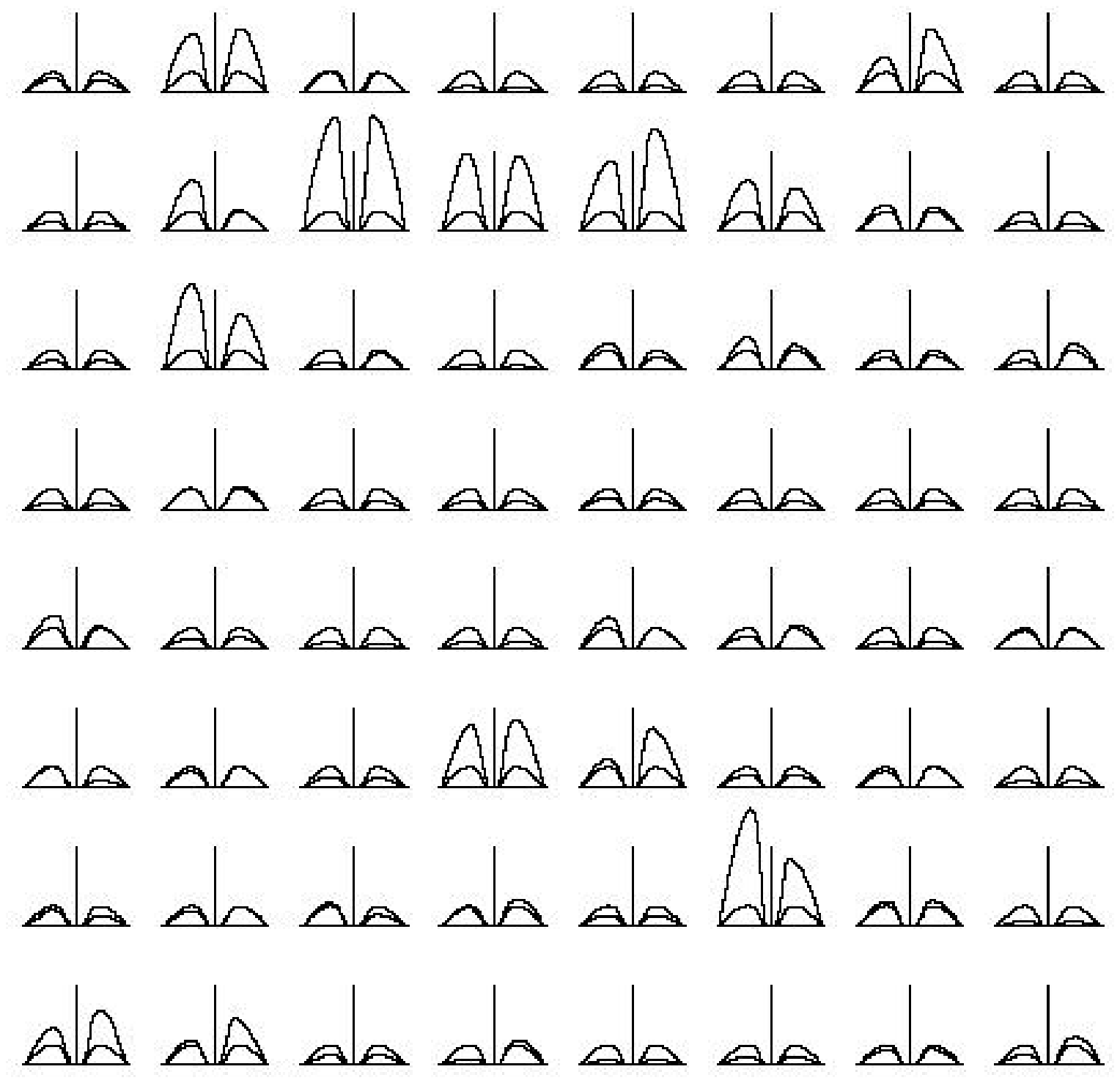}{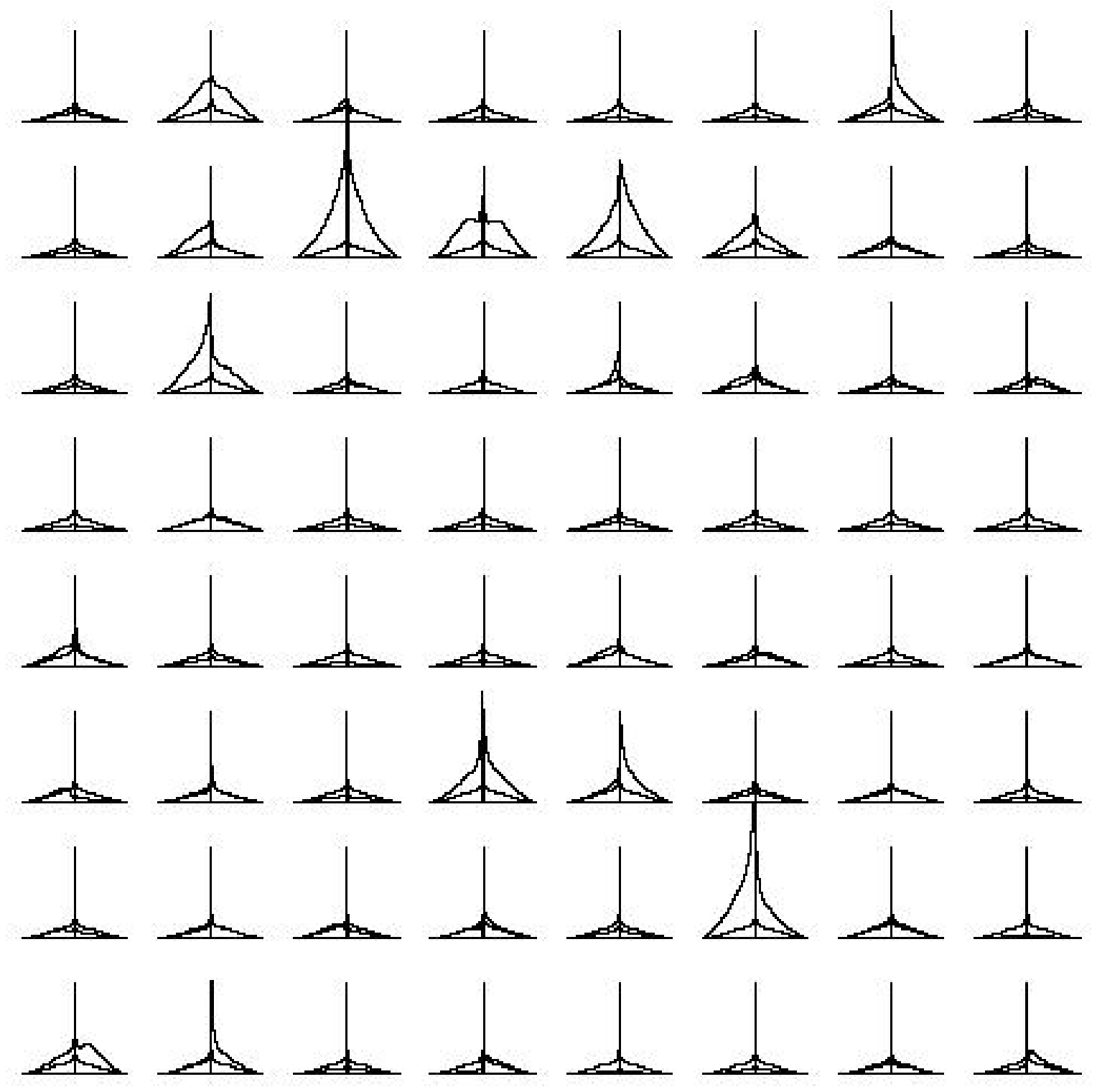}
\figcaption[]{Biconical model with $i=45^{\circ}$, $\theta_c=30^\circ$, $p=0.5$ for the left panels and $p=2$ for the right panels. The perpendicular projections of the velocity correspond to the upper panels, the parallel projections correspond to the lower panels. The heavy solid line is the amplified line profile and the lighter solid line is the line profile normalized to mean amplification in the magnification map (Fig.~\ref{fig1}).\label{fig5}}
\end{figure}

In Fig.~\ref{fig5} the bicone has an inclination of $i=45^\circ$. For the limiting geometries considered before ($i=0^\circ$ and $i=90^\circ$) there are no asymmetries induced by the microlensing in the line profile (Figs \ref{fig3} and \ref{fig4}), assuming that obscuration and optical depth effects are negligible (see Murray et al. 1995). However, for any intermediate inclination microlensing induces relative enhancements between the blue and red parts of the profile depending on the position over the magnification map. It is even possible to obtain a line profile with one part of the line amplified and other part deamplified. Notice that microlensing effects also depend on the orientation of the projected cone axis with respect to the shear.

\begin{figure}
\figurenum{6}
\plottwo{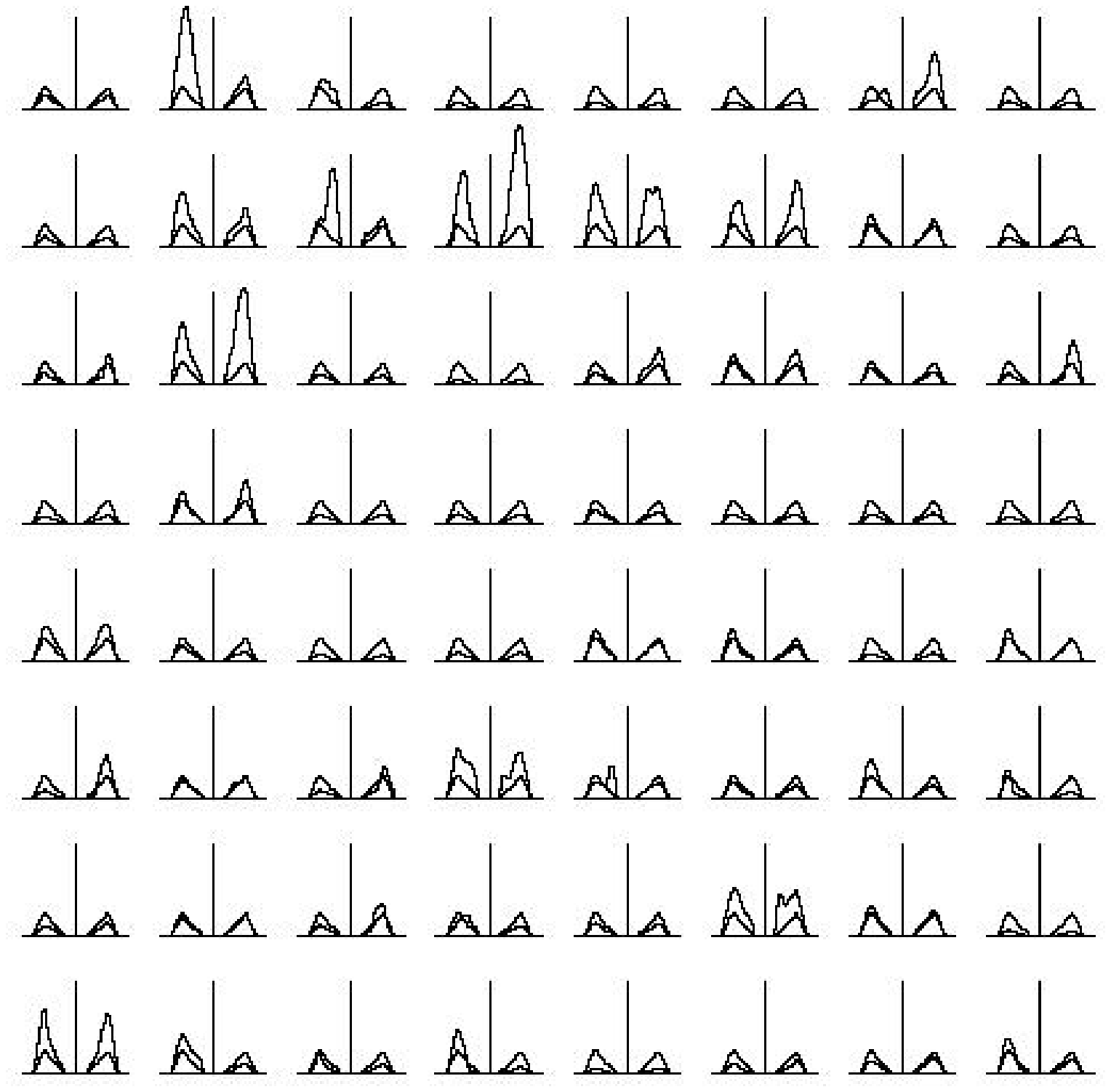}{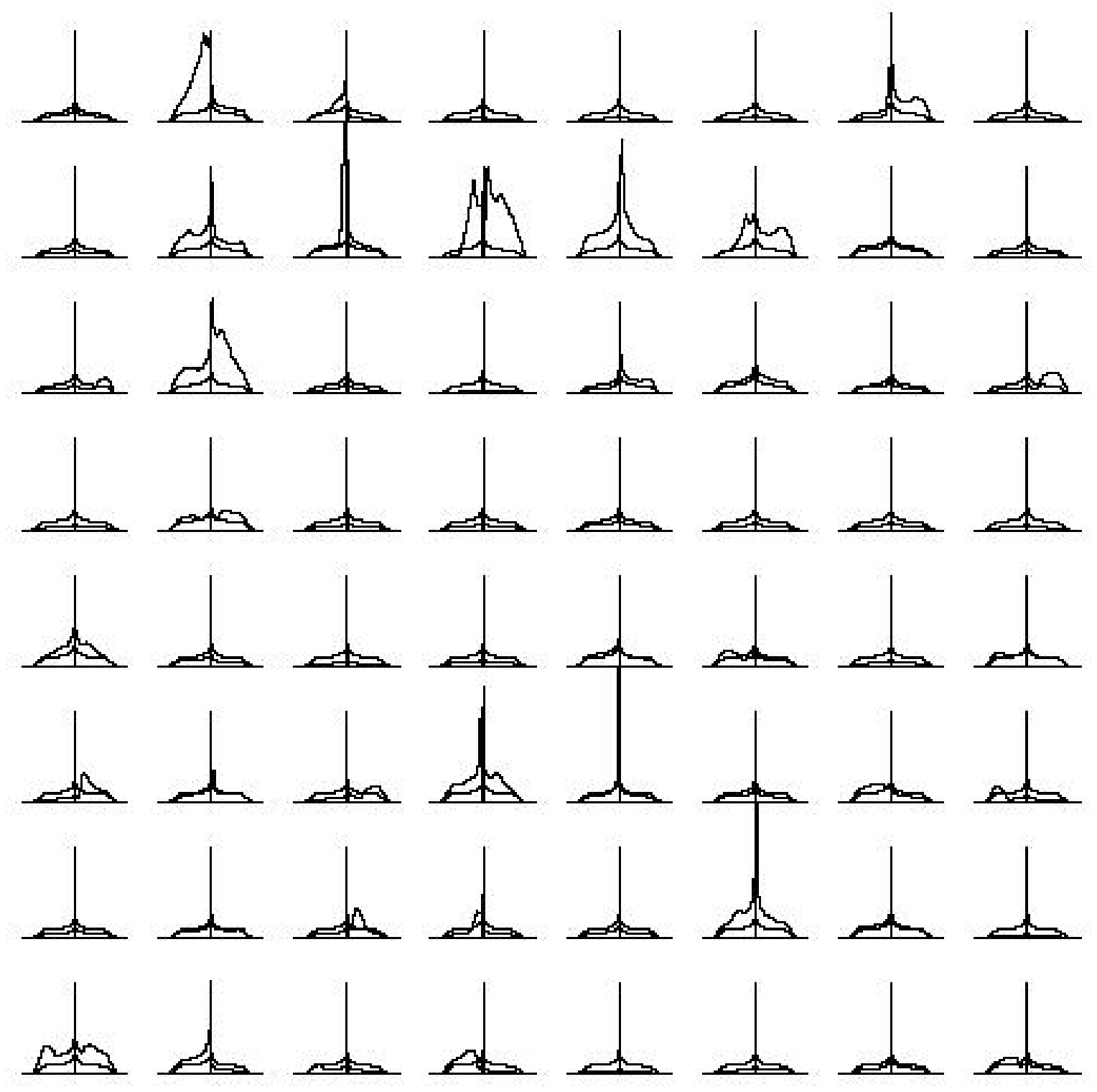}

\plottwo{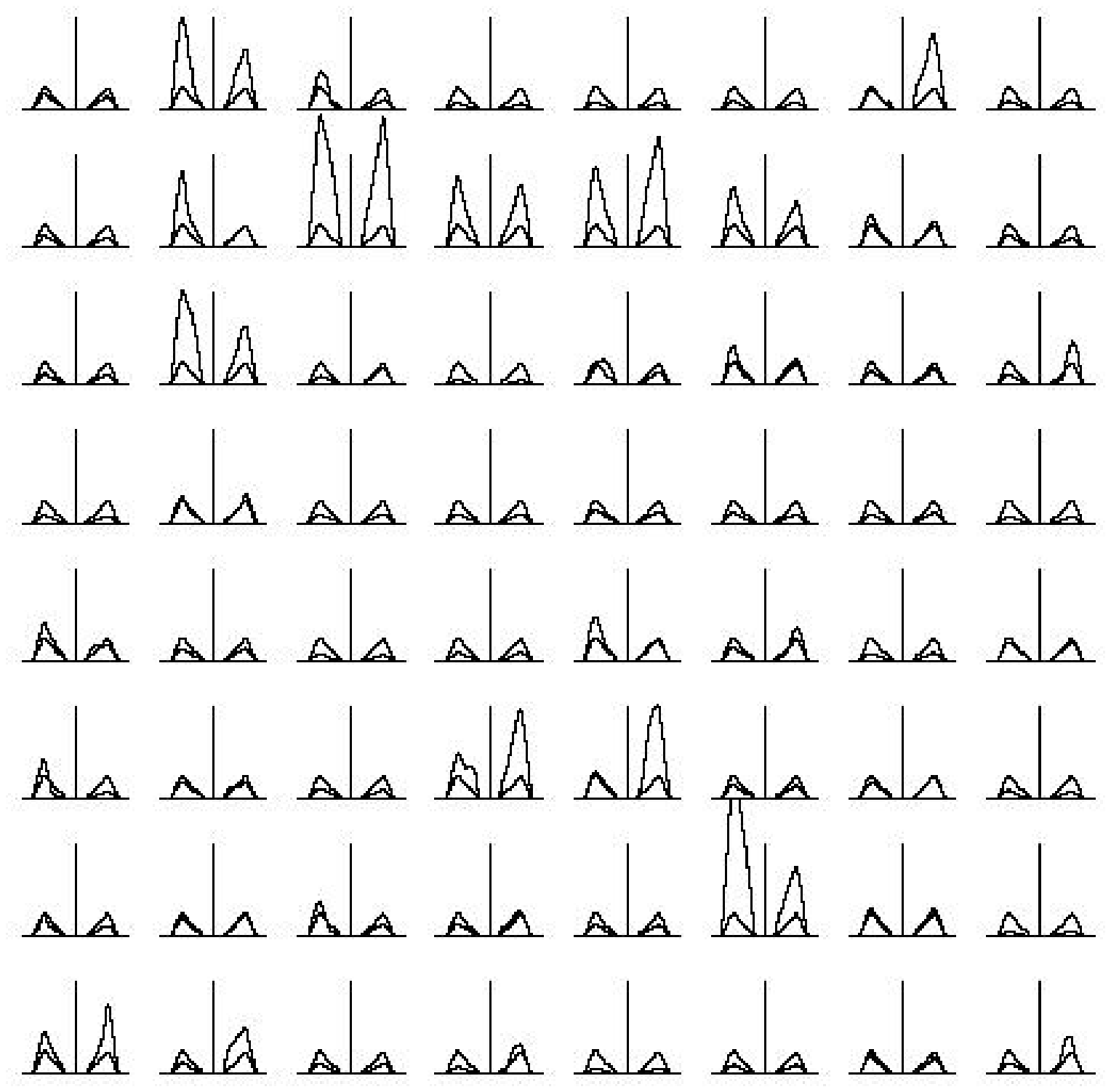}{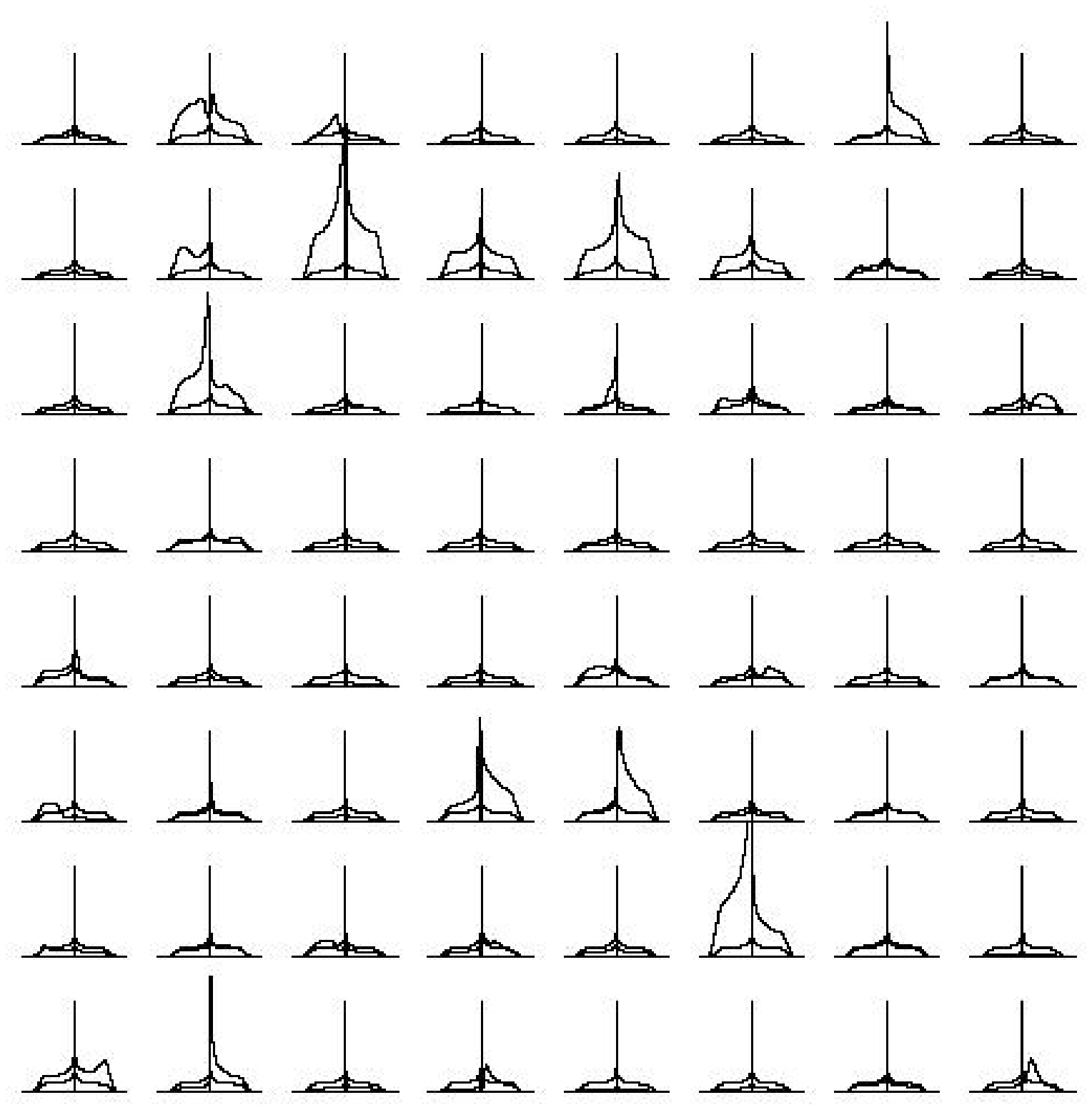}
\figcaption[]{Same as in Fig.~\ref{fig5}, but for $\theta_c=10^\circ$. \label{fig6}}
\end{figure}

In the previous cases we have considered a rather wide cone aperture ($2\cdot \theta_{\rm c}=60^\circ$). In the case of a narrower cone aperture ($2\cdot \theta_{\rm c}=20^\circ$; see Fig.~\ref{fig6}) the microlensing effects on the line profile can become extreme.

\begin{figure}
\figurenum{7}
\plottwo{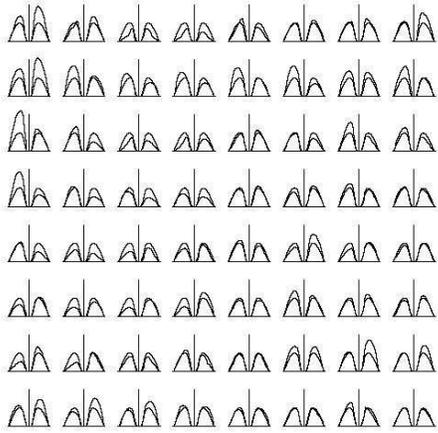}{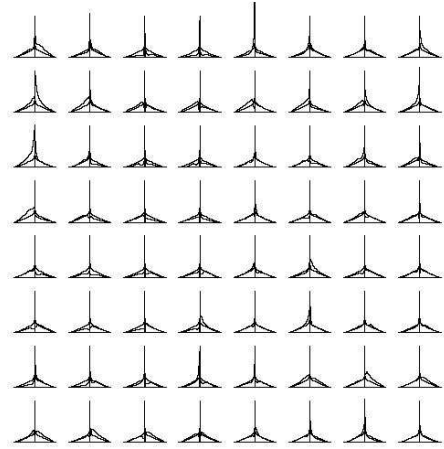}

\plottwo{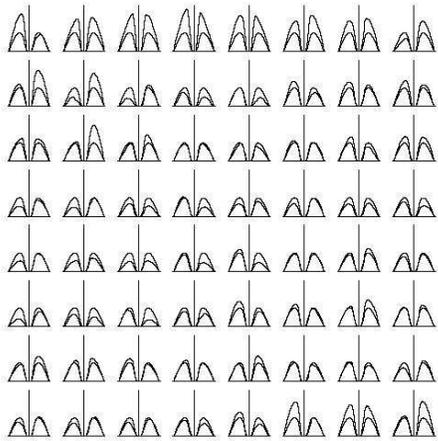}{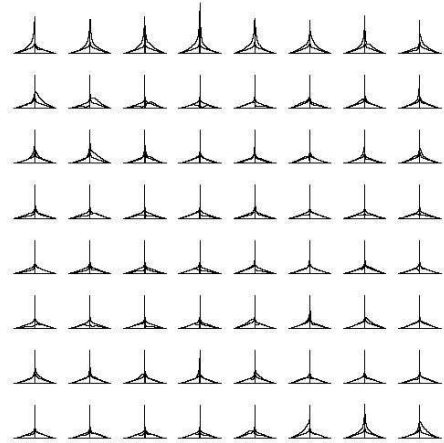}
\figcaption[]{Same as in Fig.~\ref{fig5}, but for low ionization lines ($r_{\rm ex}=3.15 \eta_0$).\label{fig7}}
\end{figure}

For the cases $i=45^{\circ}$, $\theta_c=30^\circ$, $p=0.5, 2.0$ and both bicone projections respect to the shear, we have also computed the microlensed line profiles for low ionization lines (corresponding to a $r_{\rm BLR}$=42.02 l-days, which means a $r_{\rm ex}=3.15\eta_0$). They are shown in Figure \ref{fig7}. The effects are less important but still could be detected at some positions.

\subsection{Statistics of line profiles parameters}

To obtain some statistical results we have computed emission line profiles corresponding to the cases $i=0^\circ,45^\circ,90^\circ$, $\theta_{\rm c}=30^\circ$ and both bicone projections at the position of each one of the pixels of the magnification map of the B1600+434B image (Fig.~\ref{fig1}). For each line profile we have estimated the magnification, the centroid displacement, an asymmetry parameter and the FWHM defined respectively as
\begin{equation}
\mu_{rel}=\frac{\sum_i~f_i\Delta x}{\sum_i~g_i\Delta x}
\end{equation}
\begin{equation}
C=\frac{\sum_i~x_if_i\Delta x}{\sum_i~f_i\Delta x}
\end{equation}
\begin{equation}
A=\frac{|\sum_i~f_{i+}-\sum_i~f_{i-}|}{\sum_i~f_{i+}+\sum_i~f_{i-}},
\end{equation}
\begin{equation}
FWHM=2.35\sqrt{\frac{\sum_i~x_i^2f_i\Delta x}{\sum_i~f_i\Delta x}},
\end{equation}
where $f_i$ is the amplified line profile and $g_i$ is the non-microlensed profile multiplied by the mean amplification of the map $<\mu>$. We also define the relative FWHM as $(FWHM)_{\rm rel}=FWHM(f_i)/FWHM(g_i)$

\begin{figure}
\figurenum{8}
\plotone{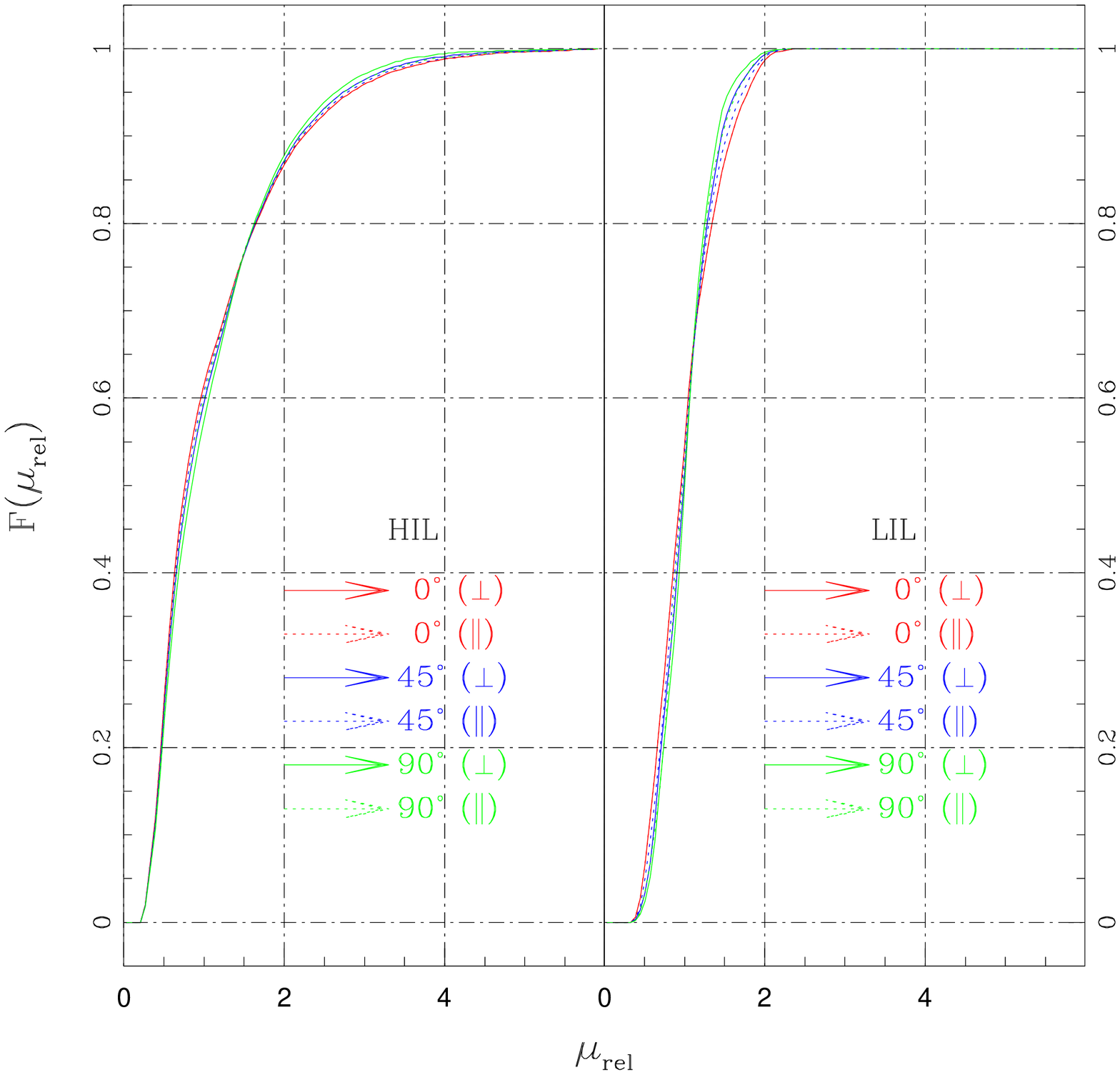}
\figcaption[]{Accumulated probability of relative magnification (see text) in B1600+434B for a biconical model. Left panel: high ionization lines ($r_{\rm ex}=0.37 \eta_0$). Right panel: low ionization lines ($r_{\rm ex}=3.15 \eta_0$). Red,  blue and green colors correspond to $i=0^{\circ},45^{\circ}$ and $90^{\circ}$, respectively. Solid and dotted lines correspond to the perpendicular and parallel projections to the shear, respectively. \label{fig8}}
\end{figure}

\begin{figure}
\figurenum{9}
\plotone{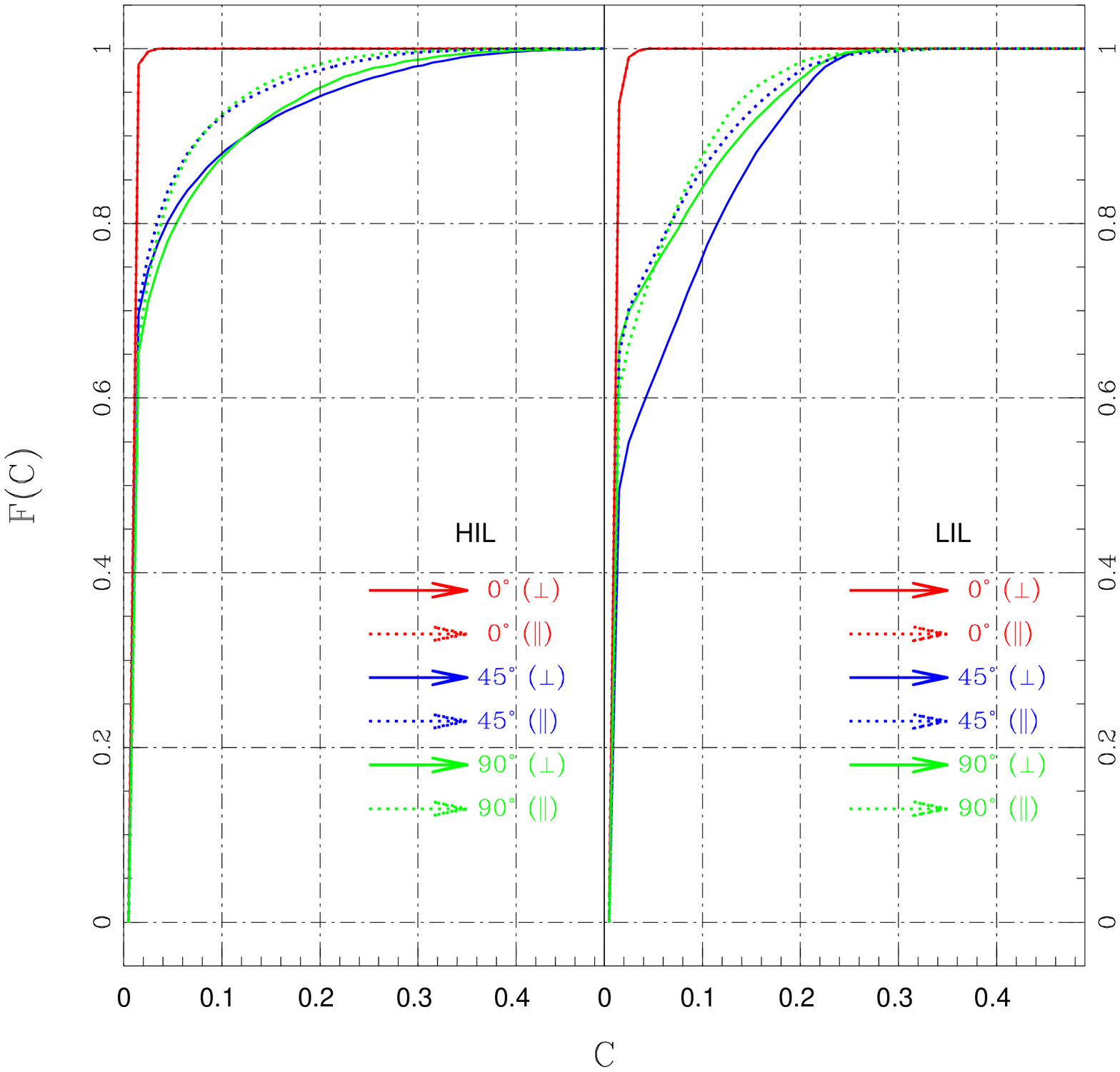}
\figcaption[]{Accumulated probability of centroid displacement (see text) in B1600+434B for a biconical model. Left panel: high ionization lines ($r_{\rm ex}=0.37 \eta_0$). Right panel: low ionization lines ($r_{\rm ex}=3.15 \eta_0$). Red,  blue and green colors correspond to $i=0^{\circ},45^{\circ}$ and $90^{\circ}$, respectively. Solid and dotted lines correspond to the perpendicular and parallel projections to the shear, respectively. \label{fig9}}
\end{figure}

\begin{figure}
\figurenum{10}
\plotone{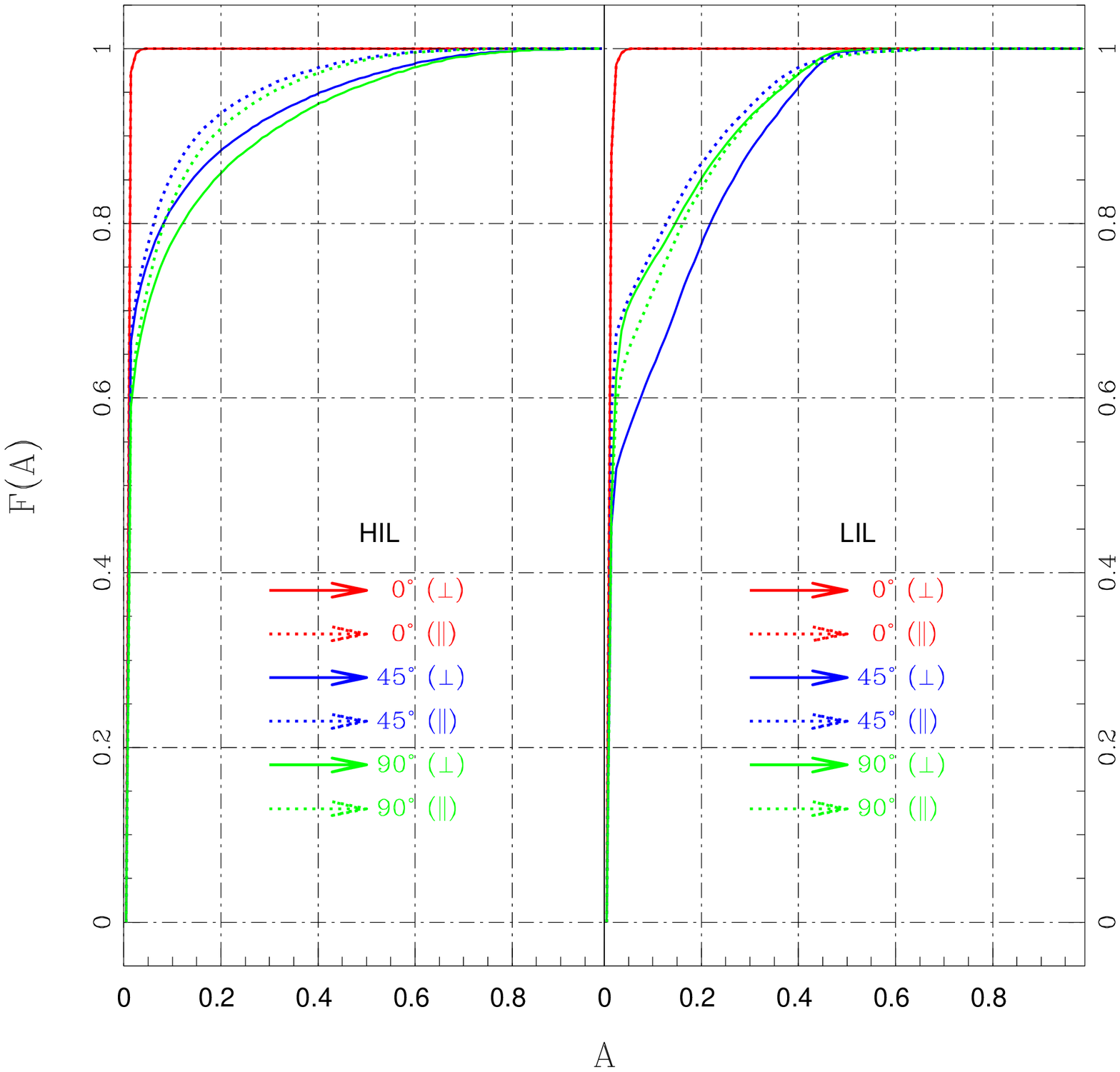}
\figcaption[]{Accumulated probability of the asymmetry parameter (see text) in B1600+434B for a biconical model. Left panel: high ionization lines ($r_{\rm ex}=0.37 \eta_0$). Right panel: low ionization lines ($r_{\rm ex}=3.15 \eta_0$). Red,  blue and green colors correspond to $i=0^{\circ},45^{\circ}$ and $90^{\circ}$, respectively. Solid and dotted lines correspond to the perpendicular and parallel projections to the shear, respectively. \label{fig10}}
\end{figure}

\begin{figure}
\figurenum{11}
\plotone{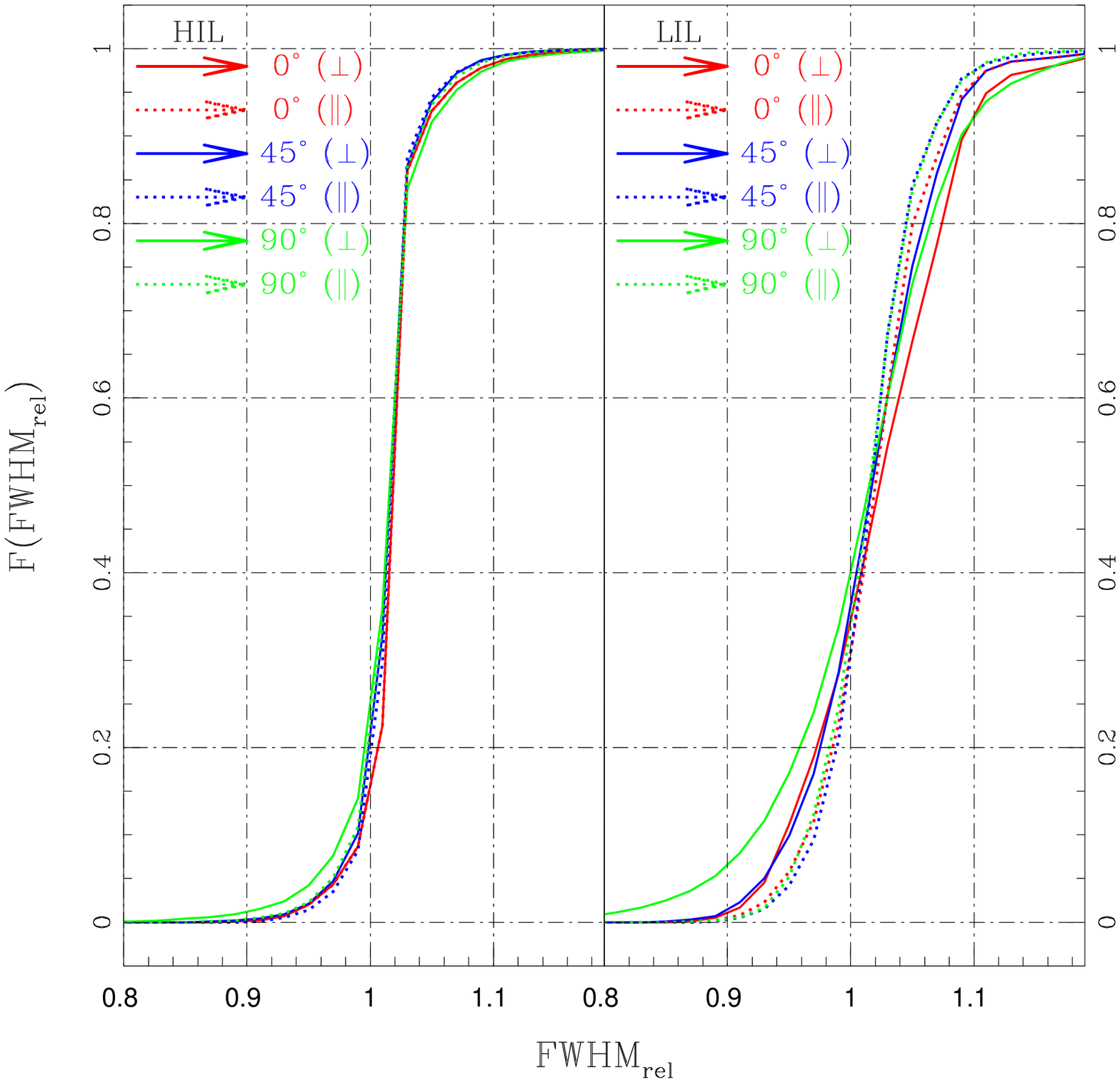}
\figcaption[]{Accumulated probability of relative FWHM (see text) in B1600+434B for a biconical model. Left panel: high ionization lines ($r_{\rm ex}=0.37 \eta_0$). Right panel: low ionization lines ($r_{\rm ex}=3.15 \eta_0$). Red,  blue and green colors correspond to $i=0^{\circ},45^{\circ}$ and $90^{\circ}$, respectively. Solid and dotted lines correspond to the perpendicular and parallel projections to the shear, respectively. \label{fig11}}
\end{figure}

The resulting accumulated probability distributions are presented in Figures \ref{fig8}, \ref{fig9}, \ref{fig10} and \ref{fig11}. In general, microlensing has a statistically significant impact on the line profiles generated by the biconical model.  For instance, for the perpendicular to the shear orientation and intermediate inclination ($45^{\circ}$) case, the probability of having a magnification greater than 1 is 38\%, the probability of a centroid displacement of a 0.1 is 18\%, the probability of an asymmetry greater than 0.2 is 17\%. However, changes in FWHM seem difficult to detect. For the case of ($90^{\circ}$) of inclination, the effects are less significant but noticeable. As expected, the effects are very small in the ($0^{\circ}$) of inclination case.

Finally, the accumulated probabilities corresponding to the low ionization lines case (with $i=0^\circ,45^\circ,90^\circ$, $\theta_{\rm c}=30^\circ$ and both bicone projections) indicate
that the probability of observing flux variations in the low ionization lines is significantly smaller for the LIL than for the HIL (this has been observed in SDSS J1004+4112, Richards et al. 2004b, Gomez-Alvarez et al. 2006). However, centroid displacements and asymmetries could also be detected in the LIL.

\section{COMPARISON BETWEEN THE LIGHT CURVES OF THE BLR AND THE CONTINUUM EMISSION REGION}

We assume a disk geometry for the region emitting the continuum radiation.
We suppose this disk to have a uniform
thickness, $h\ll r_{\rm in}$, and that the angle between its axis and
the line of sight is $i$.

We adopt a brightness distribution for the continuum source,
\begin{equation}
\label{eccont}
I_{\rm ss}(r)= 2^{p_{\rm ss}}I_{\rm ss} \left ( 1+ \frac{r^2}{R_{\rm ss}^2}
\right )^{-p_{\rm ss}},
\end{equation}
with a power-law index $p_{\rm ss}=3/2$ and $R_{\rm ss}=r_{\rm cont} \sqrt{10^{1/(p_{\rm ss}-1)}-1}$. This power-law model represents a rough version of the exact standard model (Shalyapin et al. 2002). The continuum flux is computed from
\begin{equation}
\label{fluxcont}
F ^{\rm cont}_\lambda=\int_V I_{\rm ss}(r) ~ \delta \left [\lambda-\lambda_0
\left (1+\frac{v_{\shortparallel}}{c}\right )\right ] ~ \mu(\vec{r}) ~ dV .
\end{equation}

Recent determinations by Kochanek (2004) (see also Shalyapin et al. 2002) indicate that the size of the continuum source is of about a few light days. We will adopt a continuum radius of 0.096$\eta_0$, i.e. a radius of 6 pixels over the magnification pattern that guarantee a good sampling for the continuum. This value correspond to $r_{\rm cont}=2.07$ l-days for B1600+434.

\begin{figure}
\figurenum{12}
\epsscale{.8}
\plottwo{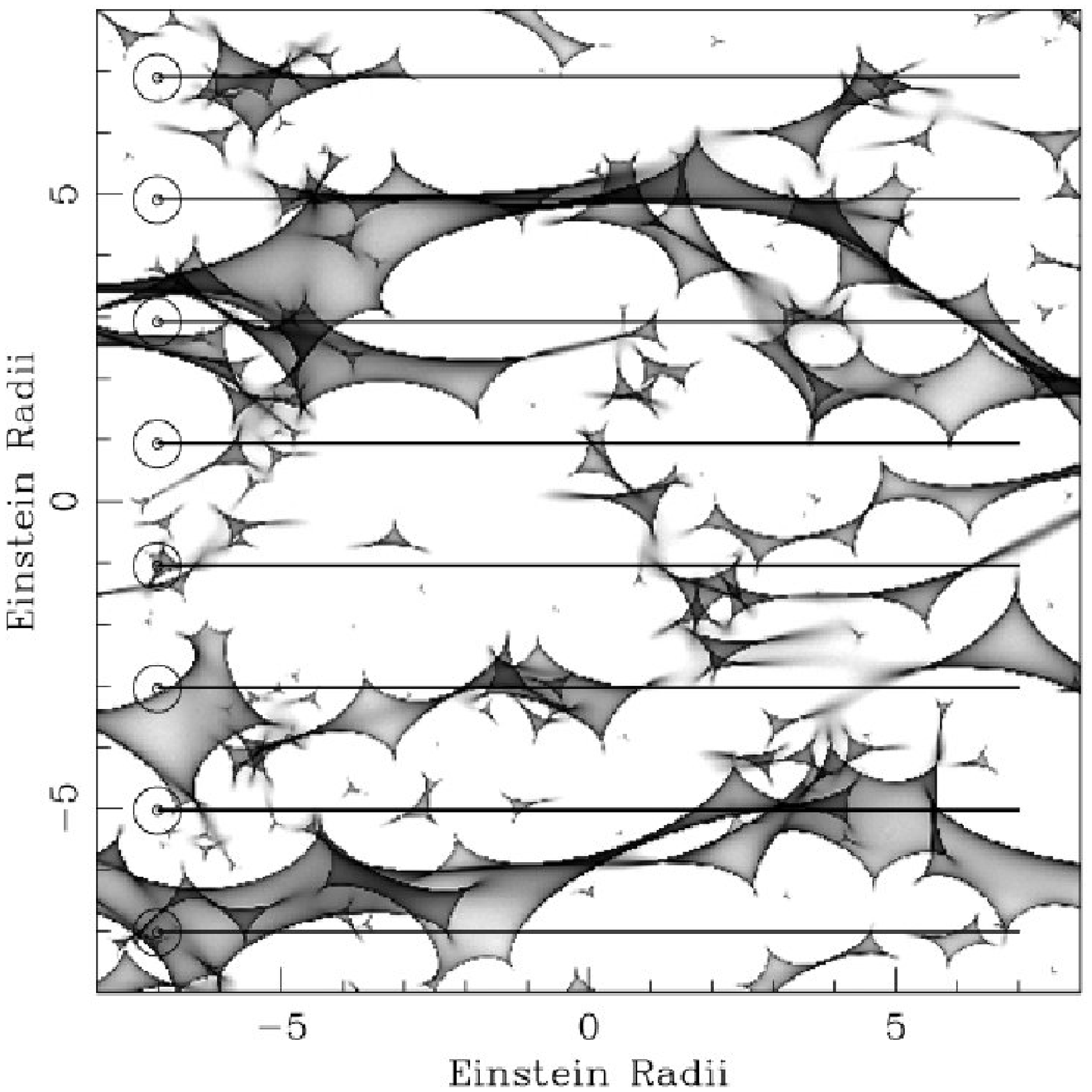}{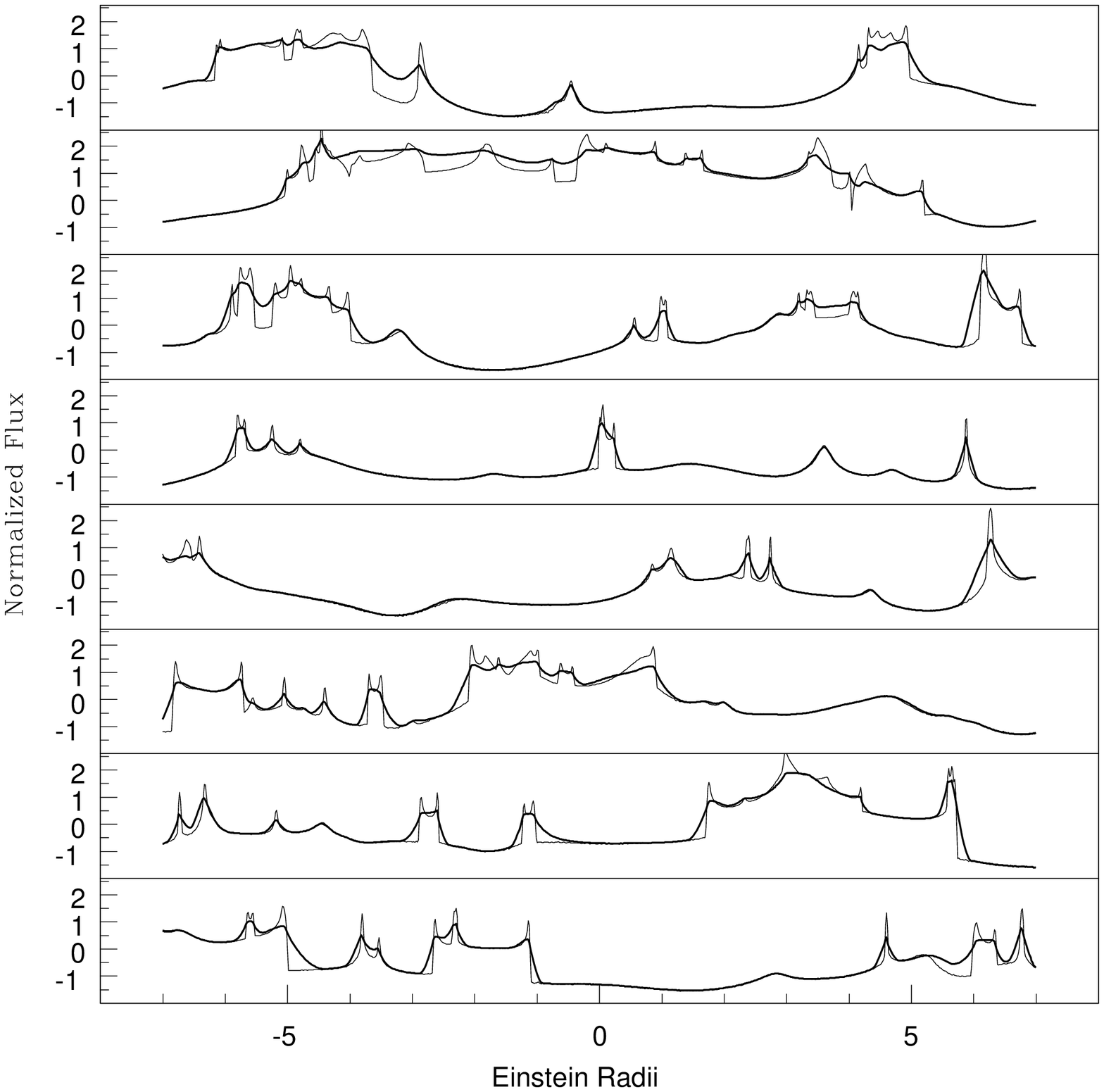}

\plottwo{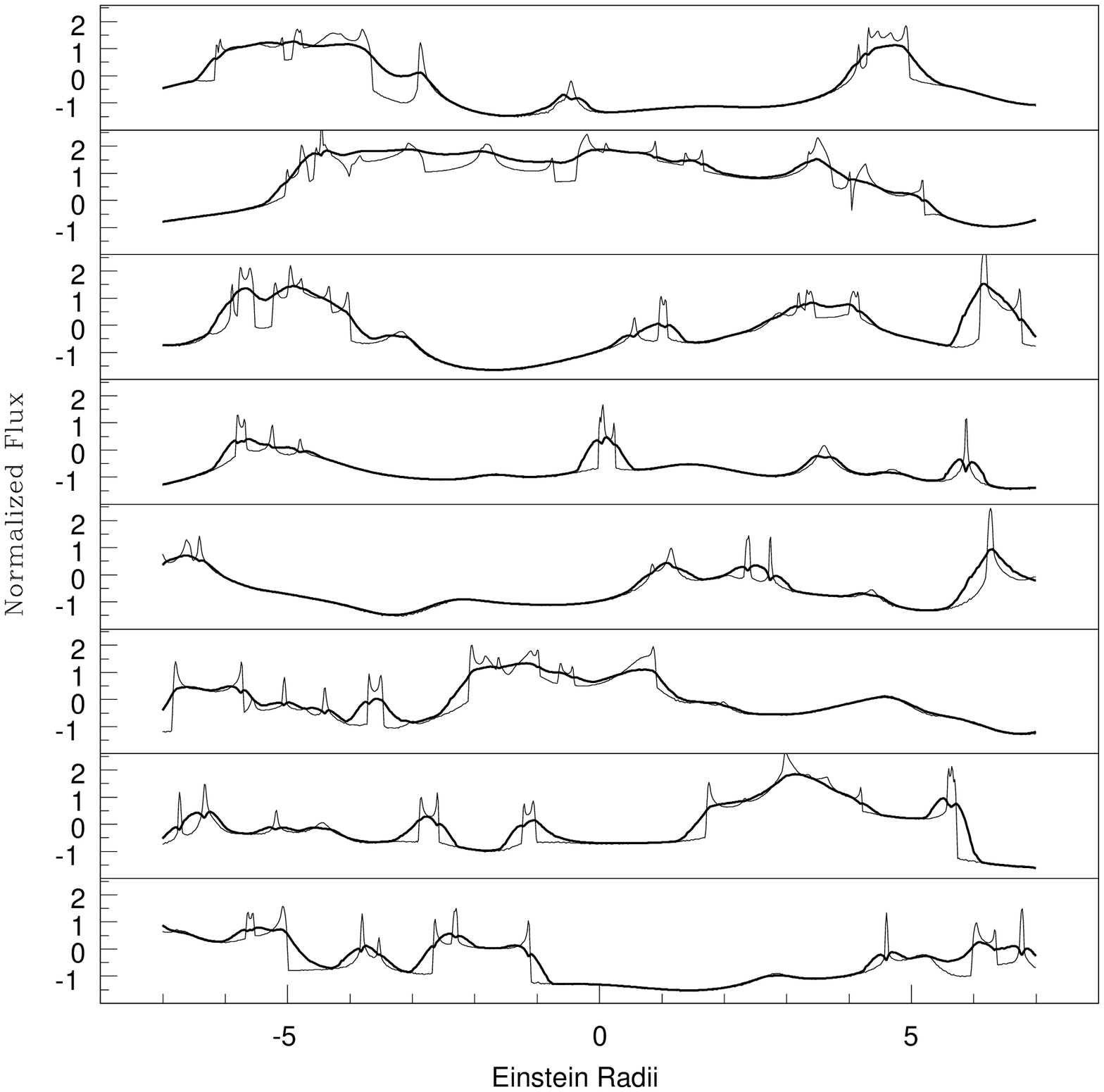}{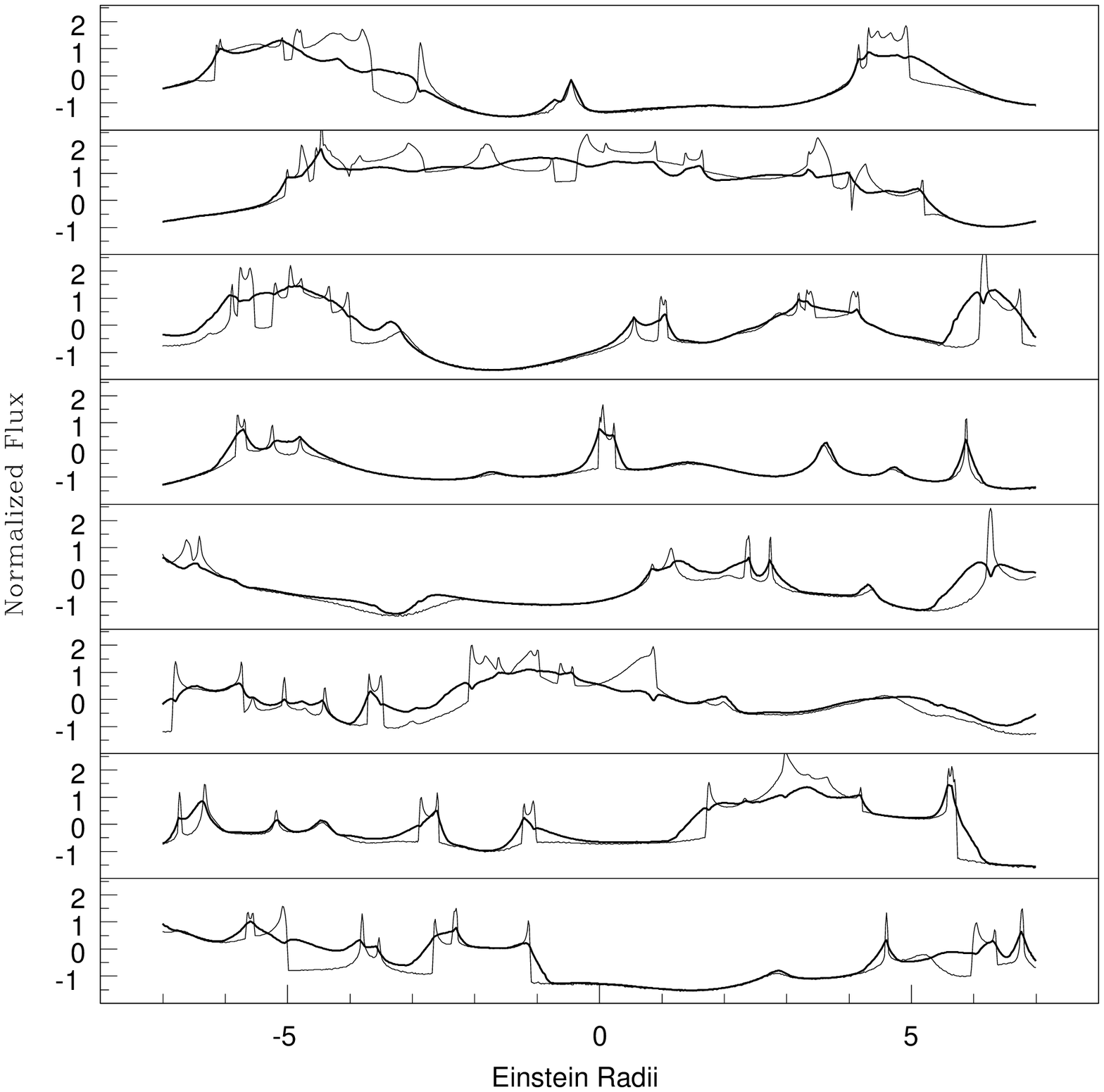}

\plottwo{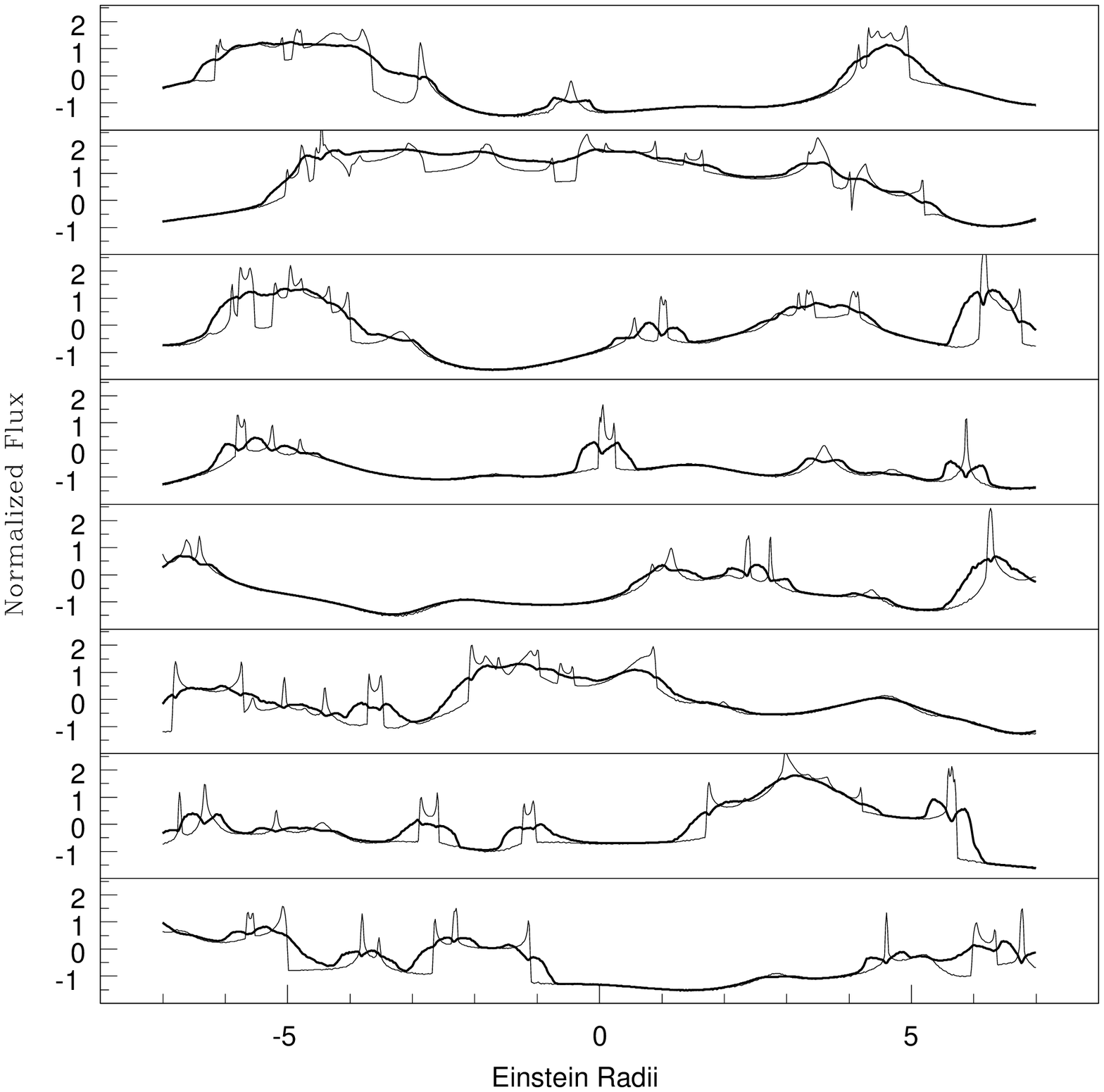}{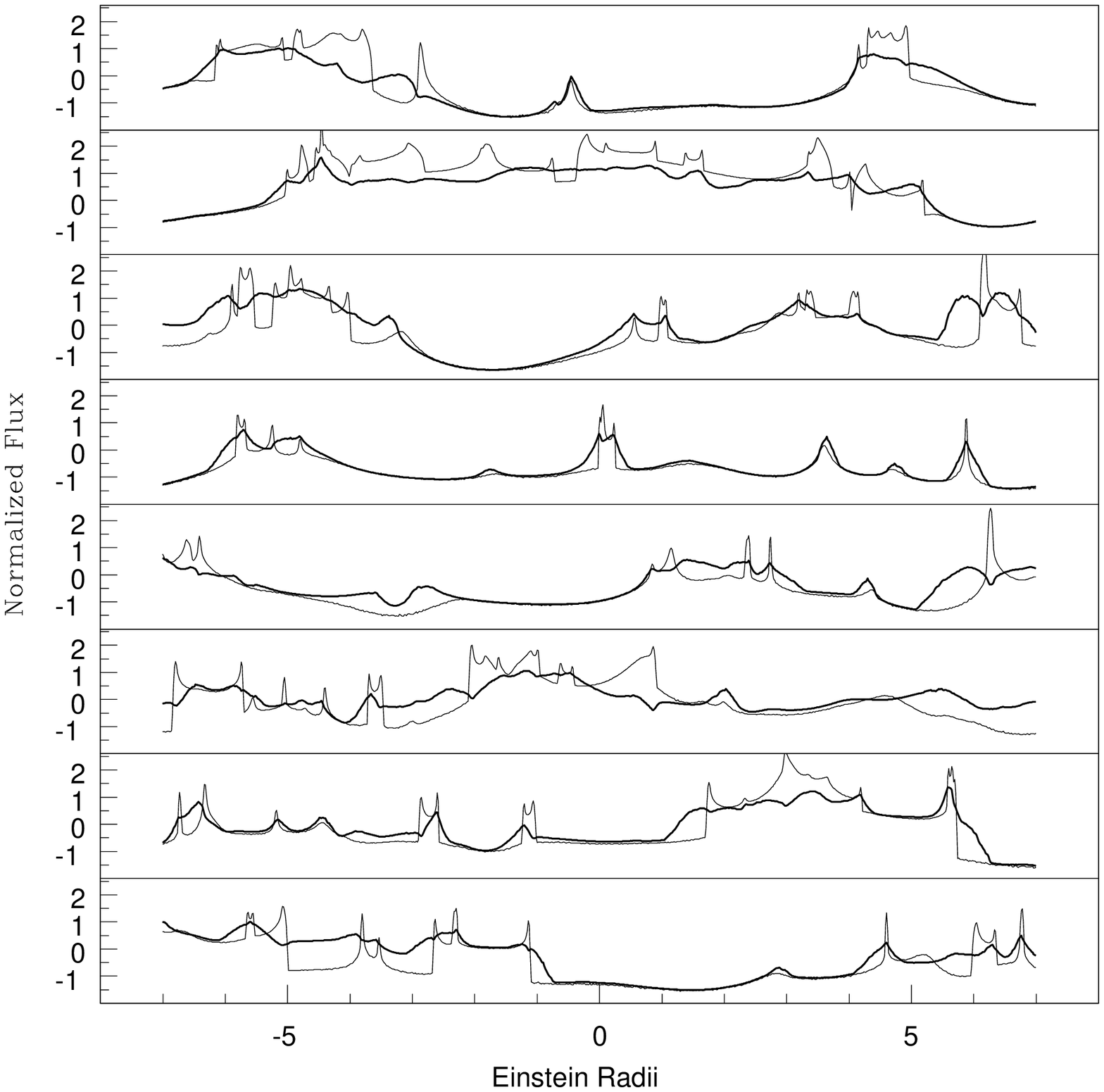}
\epsscale{1}
\figcaption{Light-curves of the BLR and the continuum represented in solid and lighter solid line, respectively. Each pair of light-curves in each panel is computed across different paths drawn in top-left panel, correspond to the magnification map in Fig.\ref{fig1}, where big circles represent to the BLR with $r_{\rm ex}=0.37\eta_0$, and small circles represent to the continuum with $r_{\rm cont}=0.096\eta_0$. The right panels correspond to the perpendicular projection of the bicone, and the left panels to parallel projection. Inclination increases from top to bottom, with $0^{\circ},45^{\circ}$ and  $90^{\circ}$. For all cases, $\theta_{\rm c}=30^{\circ}$. \label{fig12}}
\end{figure}

\begin{figure}
\figurenum{13}
\plottwo{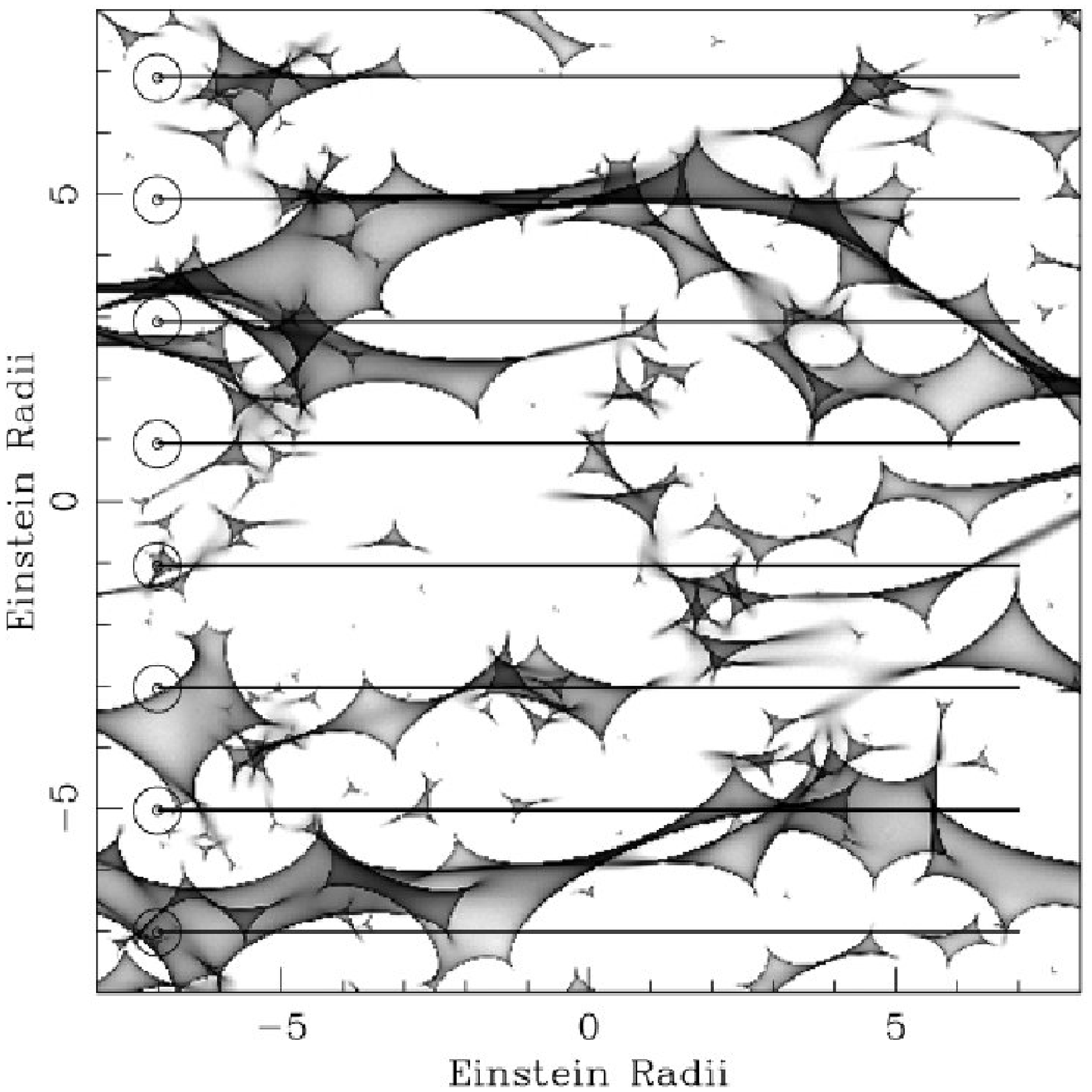}{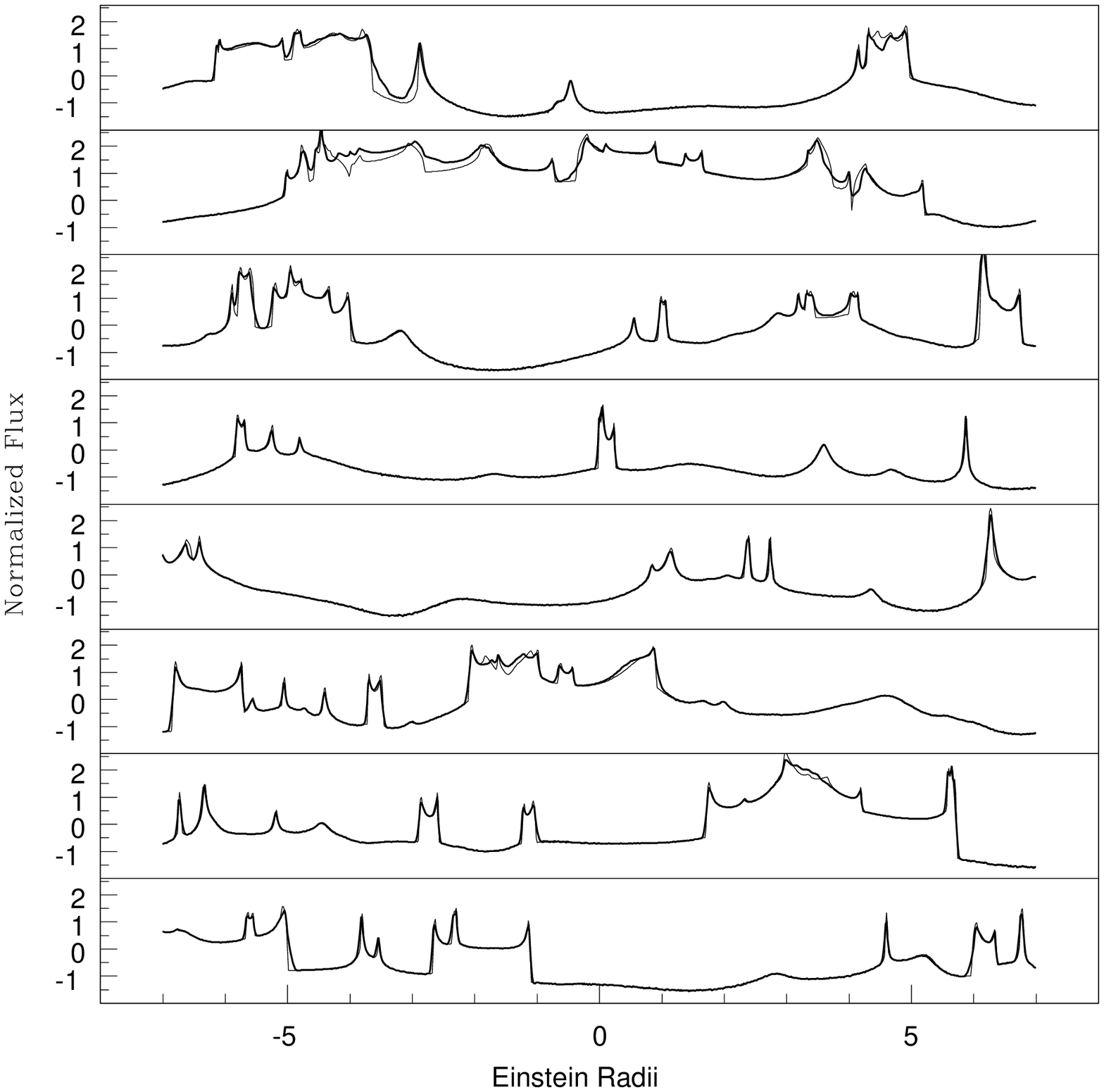}

\plottwo{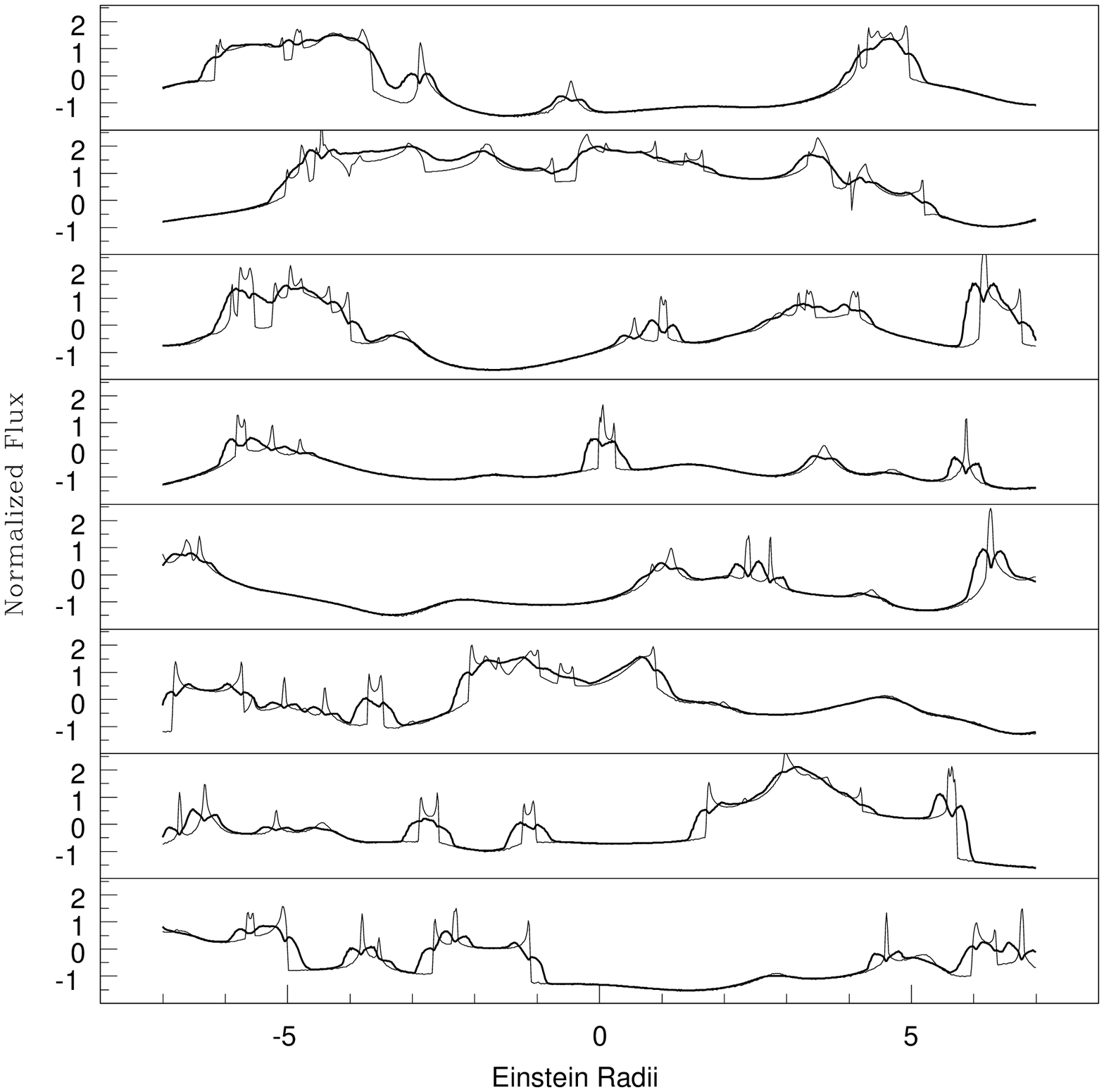}{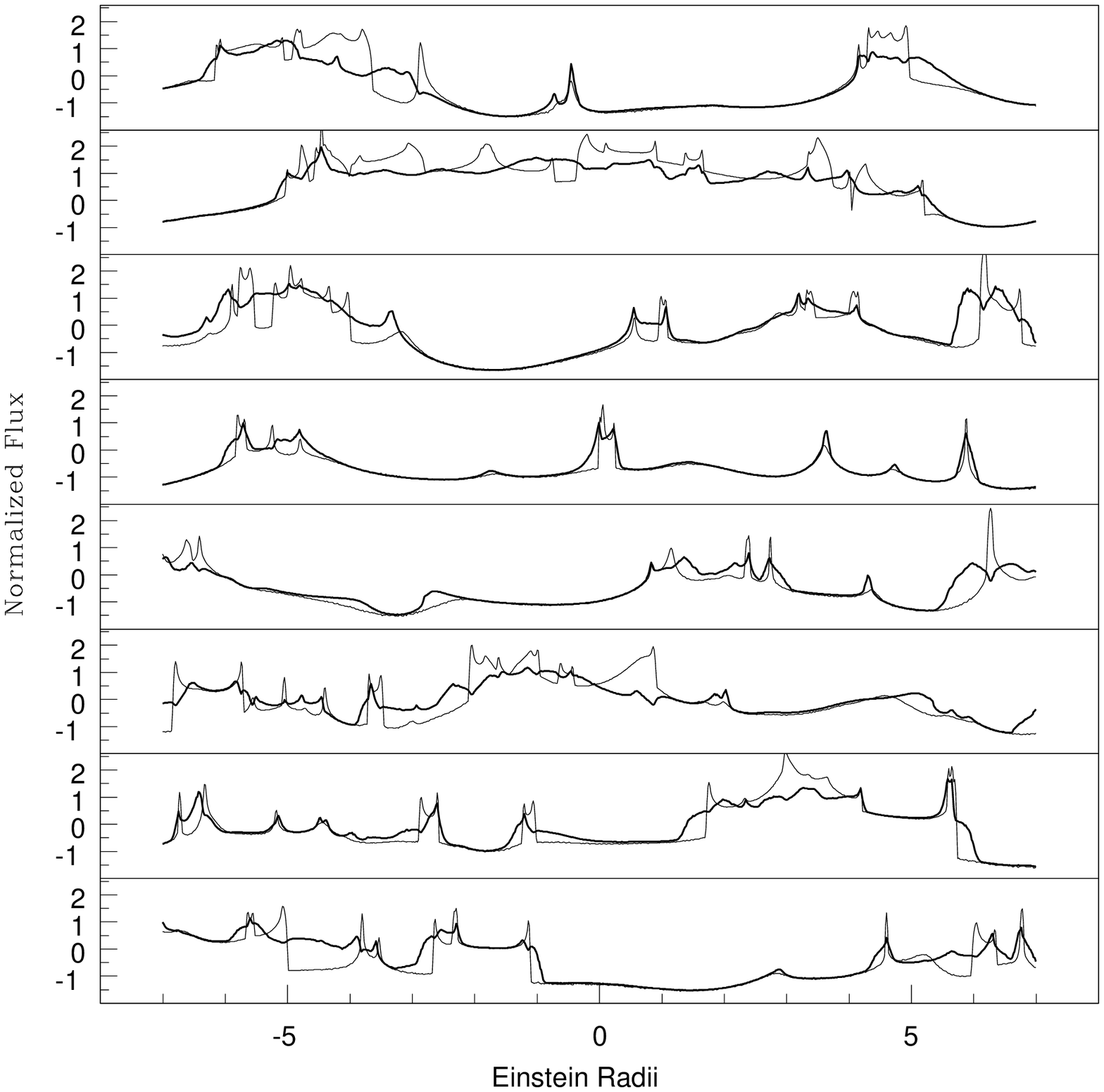}

\plottwo{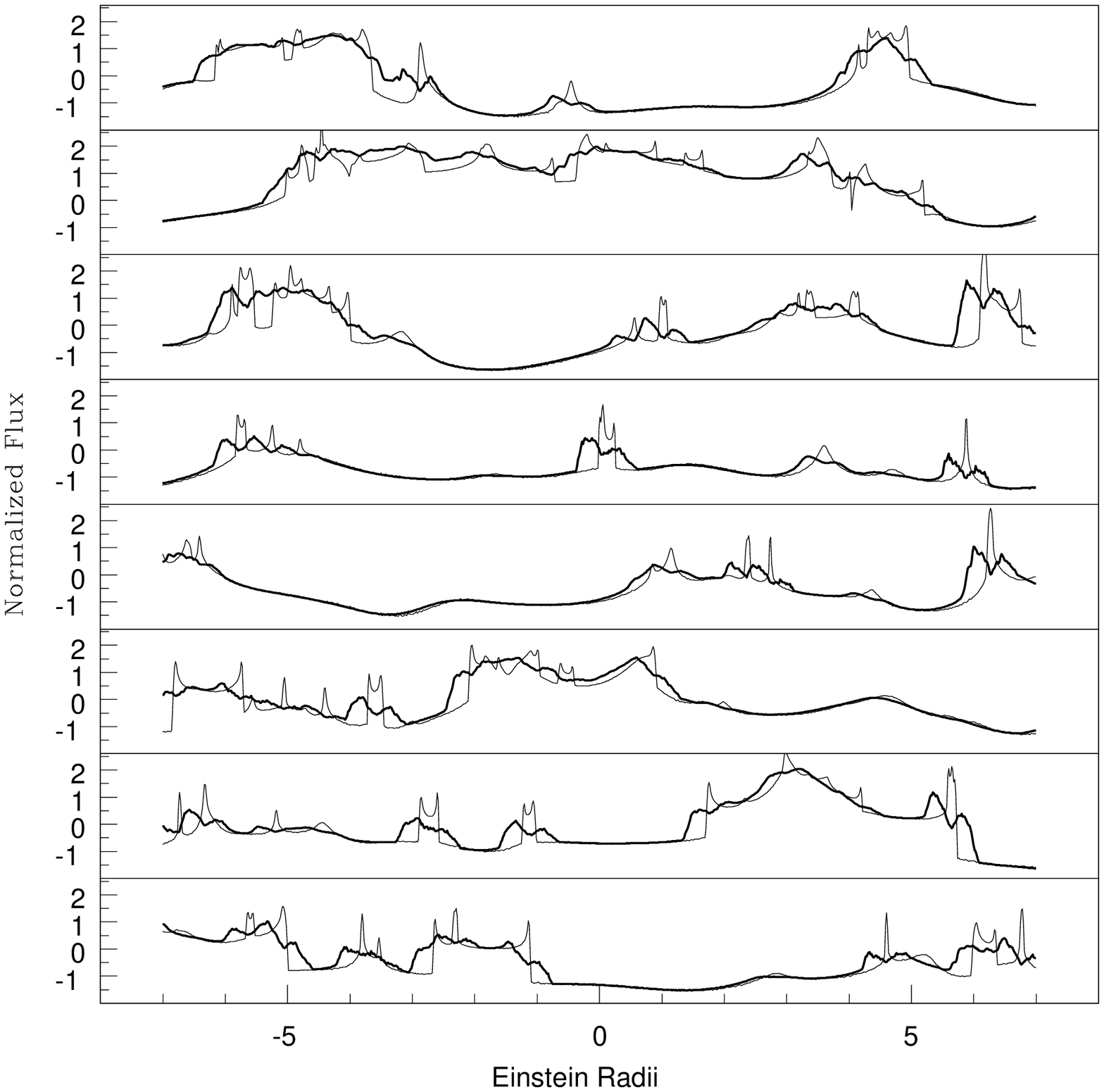}{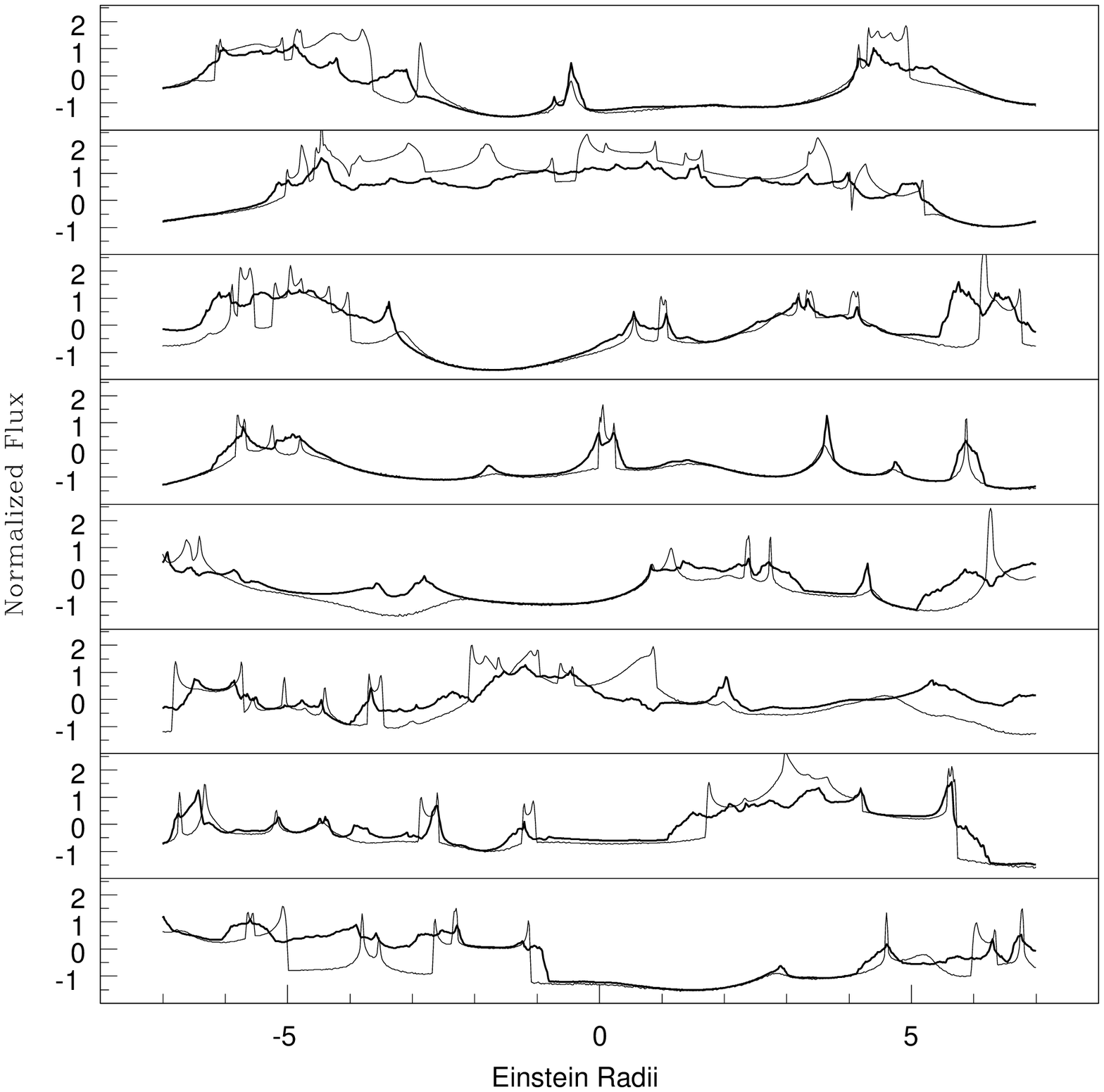}
\figcaption{Same as in Fig.~\ref{fig12}, but for $\theta_c=10^\circ$.\label{fig13}}
\end{figure}

\begin{figure}
\figurenum{14}
\plotone{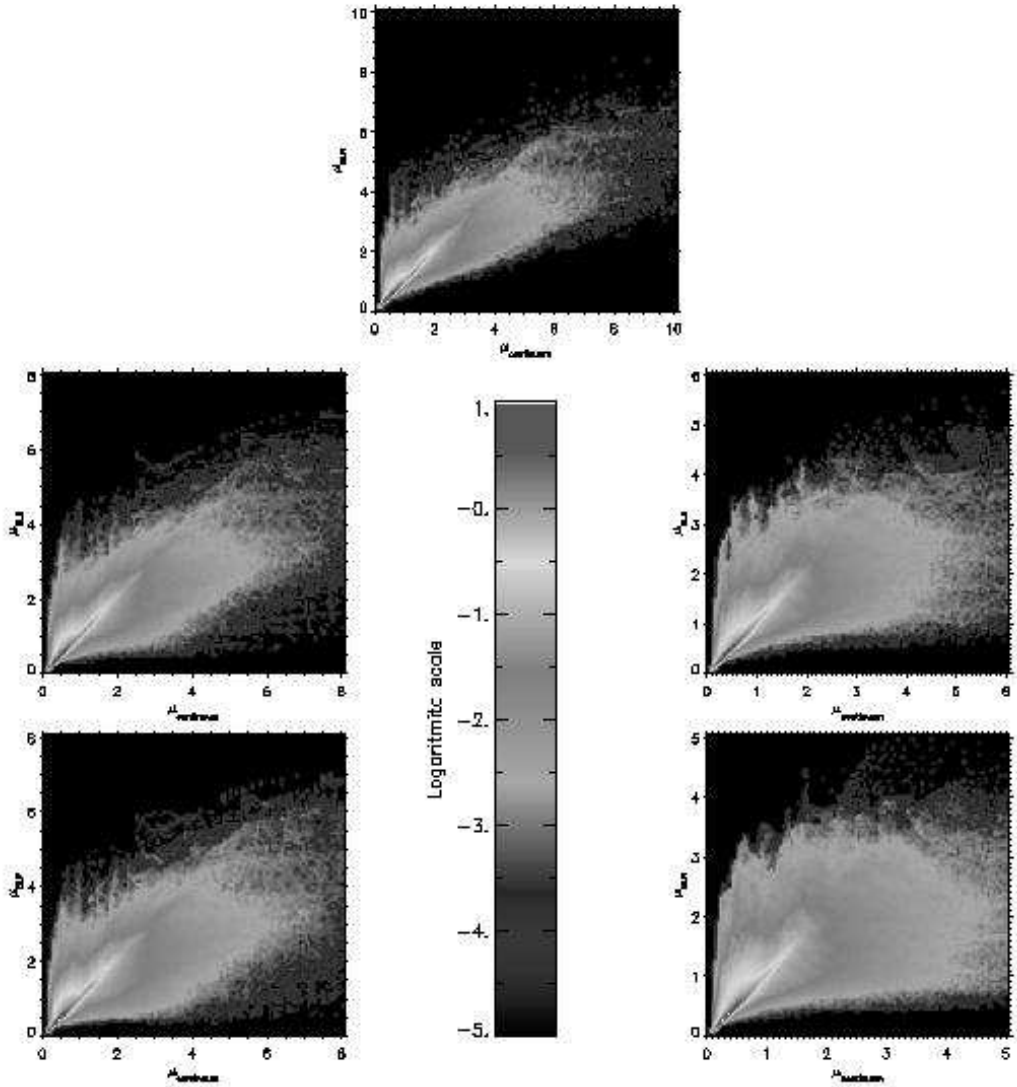}
\caption{The correlation between the amplification of the BLR and the continuum is represented, when a biconical model is considered for the BLR and a Keplerian disk for the continuum. The right panels correspond to the perpendicular projection of the bicone and the left panels to parallel projection. Inclination increases from top to bottom, with $0^{\circ},45^{\circ}$ and $90^{\circ}$. For all cases, $\theta_{\rm c}=30^{\circ}$.  \label{fig14}}
\end{figure}

\begin{figure}
\figurenum{15}
\plotone{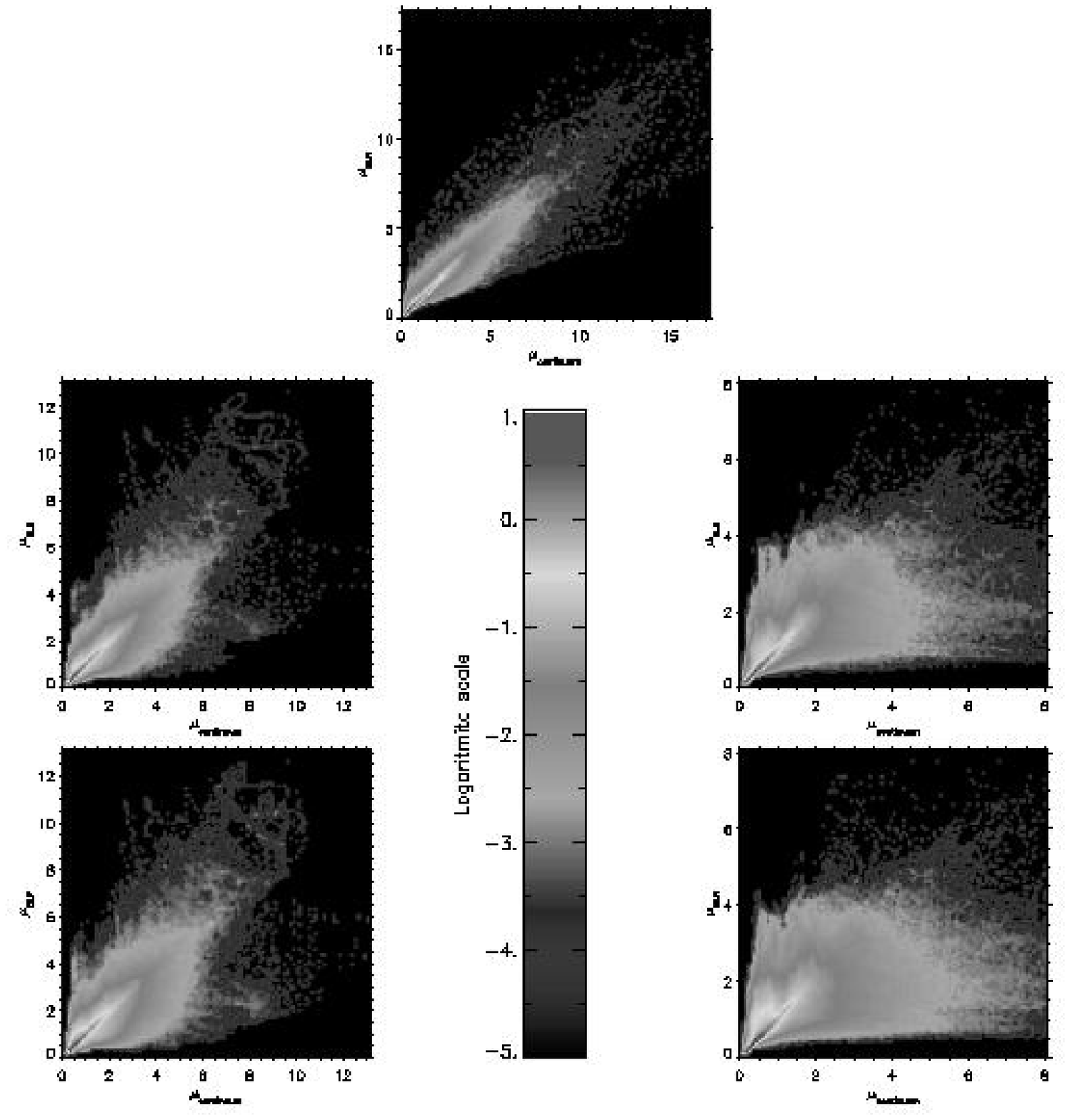}
\caption{Same as in Fig.~\ref{fig14}, but for $\theta_c=10^\circ$.\label{fig15}}
\end{figure}

The microlensing light-curve of the continuum (or the BLR) is found by convolving the brightness profile of the source with the magnification map along the path of the source over the pattern. In the top-left panel of Fig.~\ref{fig12} several paths are plotted over the magnification pattern of B1600+434B. For the half-aperture of the cones we adopt $\theta_c=30^{\circ}$. In the top-right panel the light-curves for the BLR and continuum with $i=0^{\circ}$ are shown. The middle panels correspond to  $i=45^{\circ}$ and the lower panels to $i=90^{\circ}$. Figure~\ref{fig13} is equivalent to Fig.~\ref{fig12}, but with $\theta_c=10^{\circ}$. In these figures we see that the BLR amplification may be similar to or even greater than that of the continuum (BLR amplifications two magnitudes greater than the continuum ones can be found in the perpendicular cases). For inclinations different from $i=0^{\circ}$, the degree of correlation between both the BLR and continuum light-curves depends on the orientation of the projected bicone axis with respect to the shear. The correlation is strong when the projected bicone axis is parallel to the shear direction (left panels in Figs.~\ref{fig12}, \ref{fig13}) but can be substantially reduced when it is perpendicular (right panels in Figs.~\ref{fig12}, \ref{fig13}). To gain insight into this study (see Lewis \& Ibata 2004), we have computed for each position (pixel) of the source in the magnification pattern pairs representing the BLR and continuum amplification, $(\mu_{\rm cont},\mu_{\rm BLR})$, where (from Equations \ref{fluxl} and \ref{fluxcont})

\begin{mathletters}
\begin{equation}
\mu_{\rm BLR}=\frac{\int F_{\lambda} (\mu) d\lambda}
{\int F_{\lambda} (<\mu>) d\lambda } ,
\end{equation}
\begin{equation}
\mu_{\rm cont}=\frac{\int F^{\rm cont}_{\lambda} (\mu) d\lambda}
{\int F^{\rm cont}_{\lambda} (<\mu>) d\lambda },
\end{equation}
\end{mathletters}
where $<\mu>$ is the mean magnification of the pattern. Note that the $\mu_{\rm BLR}/\mu_{\rm cont}$ ratio is related to changes in the equivalent width of the broad emission line.
Color-scale histograms representing the frequency of these pairs (Lewis \& Ibata 2004) for different values of inclination and cone aperture are represented in Figures \ref{fig14} and \ref{fig15}. In Fig. \ref{fig14} we confirm that the correlation is very strong for the case $i=0^{\circ}$, is less marked for the parallel to the shear projection of the bicone (cases  $i=45^{\circ}$, and $i=90^{\circ}$), and is weaker for the perpendicular projection (cases  $i=45^{\circ}$, and $i=90^{\circ}$). The probability of finding a microlensing event in a strip of $\mu_{\rm BLR}=\mu_{\rm cont}\pm0.5$ decreases from 78\% in the $i=0^{\circ}$ case to 65\% in the $i=90^{\circ}$ perpendicular case. In the parallel case ($i=90^{\circ}$), the probability of finding a microlensing event in this strip is  71\%. Similar results can be observed in Figure 10.

\section{IS MICROLENSING AFFECTING THE BEL IN SDSS J1004+4112?}

BEL microlensing has been tentatively detected in the quadruple large-separation
lensed quasar SDSS J1004+4112 (Inada et al. 2003) with $z_{\rm s}=1.732$ and $z_{\rm l}=0.68$. Blue-wing enhancements have
been observed in the high ionization lines in component A spectra (Richards,
Johnston \& Hennawi 2004a; G\'omez-\'Alvarez et al. 2005). A first
enhancement was detected in two epochs in May 03 and was later undetectable in
another four epochs from November to December 2003 (Richards et al. 2004b). A second
enhancement was independently detected in January 04 (G\'omez-\'Alvarez et al. 2005), March and April 2004 (Richards, Johnston \& Hennawi 2004a). In
contrast, the continuum emission has been maintained unaltered
throughout this period.

Microlensing of the high ionization BEL was the hypothesis originally preferred to explain variability (Richards et al. 2004b). However this hypothesis seems difficult to reconcile with the recurrence of the blue-wing enhancement in the framework of the typical disk-like models for the BLR: instead of a new blueward enhancement, changes of the core or the red wing of the line were expected in so far as the caustic moves across the BLR velocity field. However, this behavior would not be so strange if a biconical geometry were adopted for the BLR. According to what we have previously seen for the case $i=45^{\circ}$ and orientation of the bicone axis perpendicular to the shear, one can imagine  situations in which one of the projected cones (blue or red) could be amplified in more than one epoch without noticeable amplification of the other. In addition, the lack of correlation between the emission in the line and in the continuum is also another characteristic of the biconical model. A singular isothermal ellipsoid (SIE) model fitted to the positions of the four images of SDSS J1004+4112 reveals the presence of significant shear for image A ($\kappa=0.39$, $\gamma=0.64$).

The biconical model generates a two-peaked profile that can not realistically reproduce the emission lines in SDSS J1004+4112. However, we can consider that the bicone gives a contribution to the blue and red sides of a complex emission line. Another gaseous system will give rise to the core of the line. This two-component model has been invoked to reconcile the existence of an accretion disk (that also will generate two-peaked profiles) with AGN observations that typically show single peaked emission lines but with substructure like shoulders or bumps in the blue and red parts of the line profiles (see, e.g., Popovi{\'c} et al. 2004 and references therein).  Thus,  microlensing magnification of one of the peaks of the biconical model could produce an enhancement of one of the wings or an asymmetrical shoulder. To add more flexibility to the model we will also consider that one of the cones could be strongly obscured. Thus, our aim with this very simple model is mainly to test whether the qualitative features of the lines variability can be reproduced with microlensing of a bicone or not.

\begin{figure}
\figurenum{16}
\begin{center}
\plotone{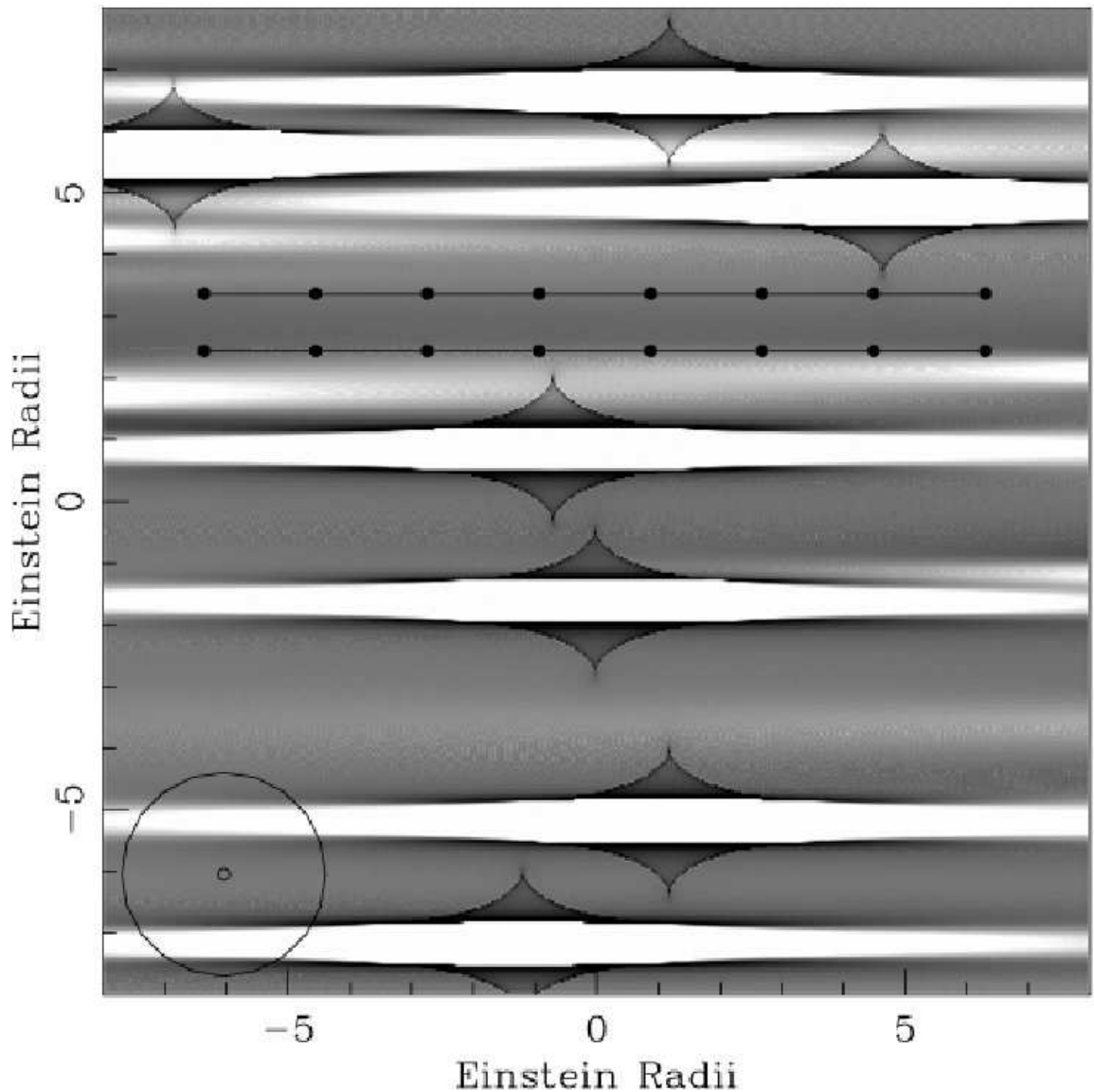}
\end{center}
\caption{The magnification map for the A image of SDSS J1004+4112 for $1\%$ of total mass in stars. Each point represents to BLR in different positions over the magnification map, and the lines are the paths over the map where the light-curves are computed. The biggest circle represents to the BLR with $r_{\rm ex}=1.65\eta_0$, and the smallest circle represents to the continuum with $r_{\rm cont}=0.096\eta_0$. \label{fig16}}
\end{figure}

\begin{figure}
\figurenum{17}
\plotone{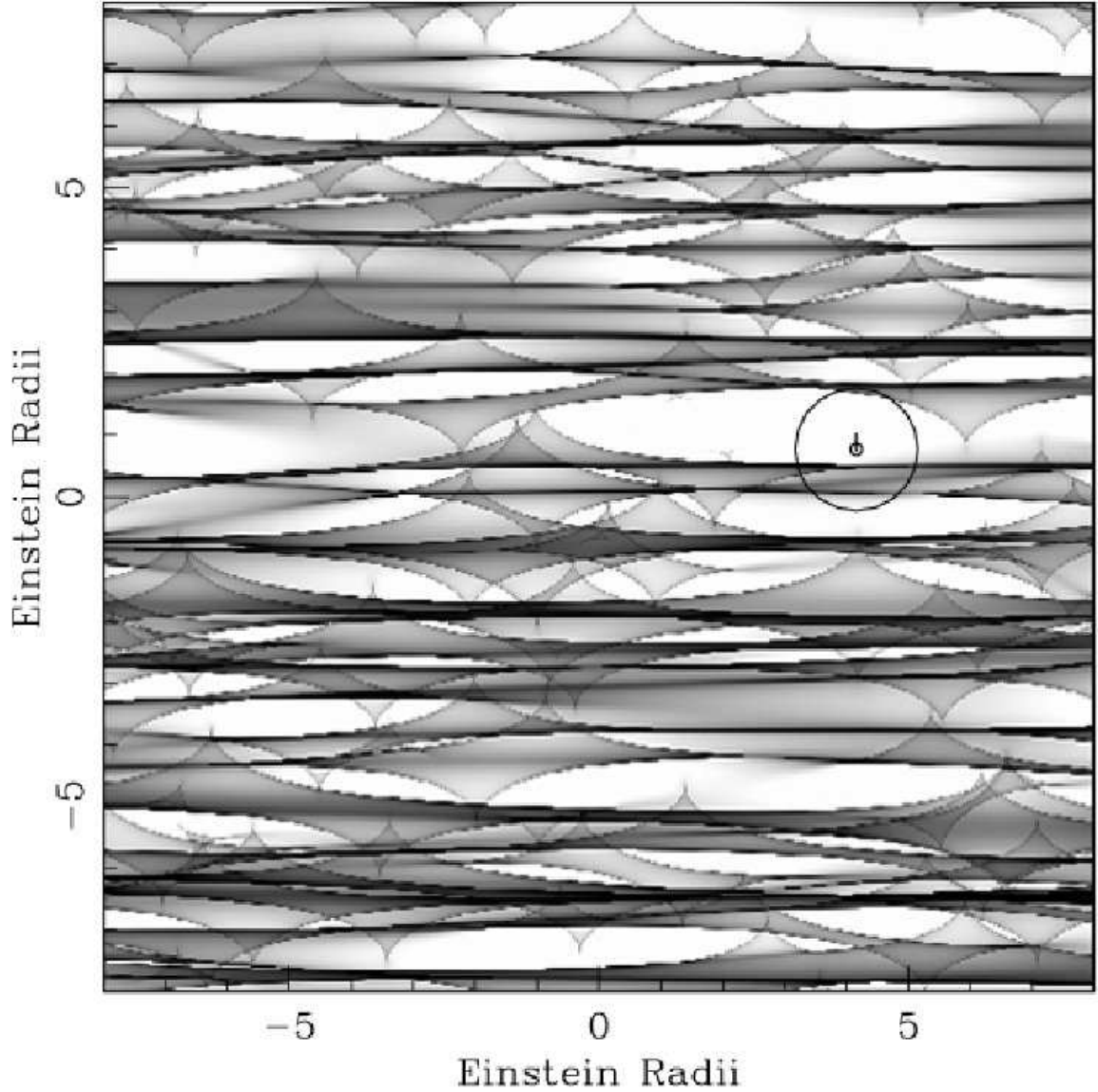}
\caption{The magnification map for the A image of SDSS J1004+4112 for $10\%$ of total mass in stars. The biggest circle represents to the BLR with $r_{\rm ex}=1\eta_0$, and the smallest circle represents to the continuum with $r_{\rm cont}=0.096\eta_0$. The small line starting from the circles centre is the path over the magnification pattern where the light-curves are computed. \label{fig17}}
\end{figure}

In Figs~\ref{fig16} and \ref{fig17} magnification maps for SDSS J1004+4112A corresponding to two different fractions of matter in stars with respect to the total mass (1\% and 10\%) are represented.
An area of 500$\xi_0\times500\xi_0$ was considered at the lens plane to compute a magnification pattern covering an area of 16$\eta_0\times16\eta_0$
($1000\times 1000$ pixels) at the source plane. 256 rays were shot per
unlensed pixel, which means $2.5\cdot10^{11}$ rays traced between the observer
and the source. The number of stars considered was 312 and 3121 for the 1\% and 10\% fractions of matter in stars, respectively. These values correspond to $\kappa_{\rm c}=0.386$ (1\%) and $\kappa_{\rm c}=0.351$ (10\%). The theoretical mean amplifications, $<\mu_{th}>=|(1-\kappa)^2-\gamma^2|^{-1}$, is 23.261 and the mean amplifications of the maps are 23.105 (1\%) and 21.611 (10\%), respectively. The most noticeable feature of both maps (Figures \ref{fig16} and \ref{fig17}) is  the characteristic pattern of alternating high and low amplification stripes induced by the strong elongation of caustics in the direction of the shear.

We consider global displacements of the BLR parallel and transversal to the shear to try to reproduce the main features of the spectra variability in J1004+4112. The BLR radius for this quadruply imaged quasar is $r_{\rm BLR}=7.57$ l-days for high ioni\-za\-tion lines (using NGC 5548 as a reference object). In this gravitational lens $1\eta_0 (1M_{\odot})=16.61$ l-days and, hence, $r_{ex}=0.74\eta_0$. As in the case of B1600+434 the continuum radius is 0.096$\eta_0$ (6 pixels over the magnification pattern) that correspond to $r_{\rm cont}=1.59$ l-days. The emissivity law is $\beta=0$.

\subsection{Parallel to the Shear Paths}

\begin{figure}
\figurenum{18}
\plotone{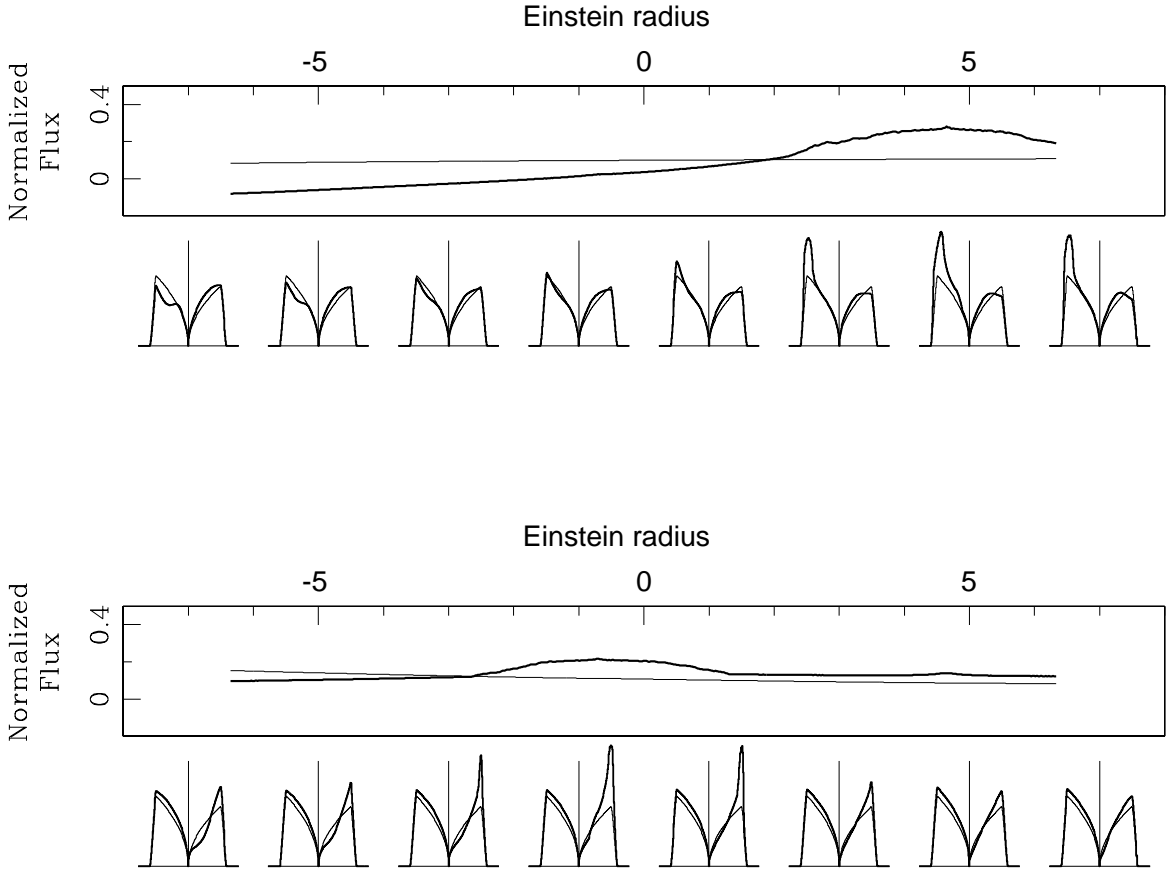}
\caption{The line profiles for the BLR for each point and the light-curves for the BLR and the continuum for each path marked in Fig. \ref{fig16} are shown. The parameter are $i=45^{\circ}$, $\theta_c=5^\circ$, $p=2$, $r_{\rm cont}=0.096\eta_0$, $r_{\rm ex}=1.65\eta_0$ and $r_{\rm in}=0.1\eta_0$. In the line profile panels the heavy solid lines represent the amplified line profile and the lighter solid lines correspond to the line profile normalized to mean amplification in the magnification map. In the
light-curve panels the solid line represents the light-curve from the BLR and the lighter solid line corresponds to the light-curve from the continuum. \label{fig18}}
\end{figure}

In this case relative enhancements of any of the line profile peaks without noticeable continuum amplification are induced in a natural way by the presence of alternating stripes of high and low amplification. We have computed BEL profiles with $r_{\rm ex}=1.65\eta_0$, which means a more conservative value for the BLR size, $r_{\rm BLR}=17$ l-days, or that the mass of stars in the lens are $0.2M_\odot$, with $1\eta_0 (0.2M_\odot)=7.43$ l-days. These profiles are convolved with the magnification pattern corresponding to a mass in stars amounting to 1\% of the total mass (Fig. \ref{fig16}), following two different tracks over the pattern. As can be seen in Fig. \ref{fig18}, the continuum remains unaltered along the entire path. However, we face here the same problem as Richards et al.\ (2004b): the enhancement event lasts several Einstein radii but according to the time estimates of Richards et al. (2004b) it should last about one tenth of an Einstein radius (for 0.1$M_\odot$). In principle, to reduce the time-scale of an event (and also the lapse between events) we could increase the number of caustics by increasing the fraction of mass in stars. However the caustic elongation in the shear direction is so strong that the time-scales of the events are not substantially reduced. The main effect of increasing the mass fraction in stars is to reduce the distance between consecutive stripes in the magnification patterns.

\subsection{Transversal to the Shear Paths}

\begin{figure}
\figurenum{19}
\plotone{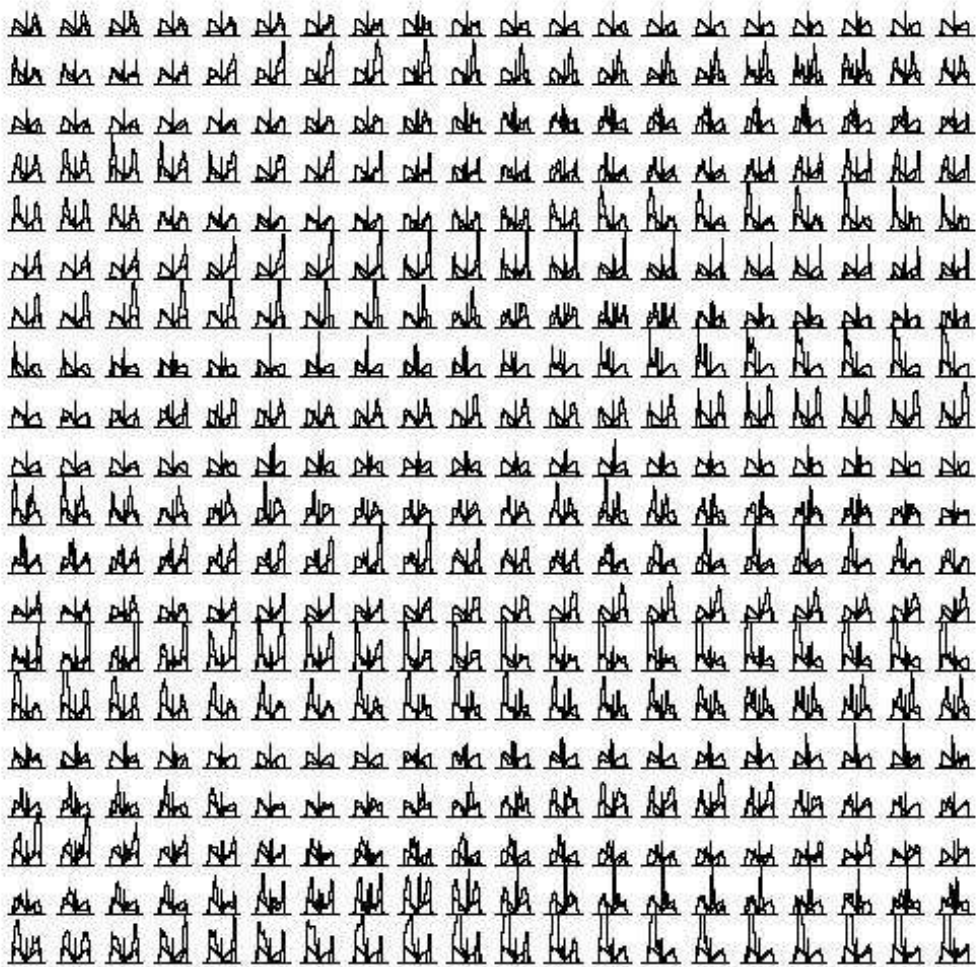}
\caption{The BEL profiles for 20 $\times$ 20 locations regularly distributed in the magnification pattern (Fig. \ref{fig17}) are shown. The parameter are  $i=45^{\circ}$, $\theta_c=5^\circ$, $p=2$, $r_{\rm ex}=1 \eta_0$ and $r_{\rm in}=0.1\eta_0$. The heavy solid lines represent the amplified line profile and the lighter solid lines correspond to the line profile normalized to mean amplification in the magnification map.\label{fig19}}
\end{figure}

\begin{figure}
\figurenum{20}
\plotone{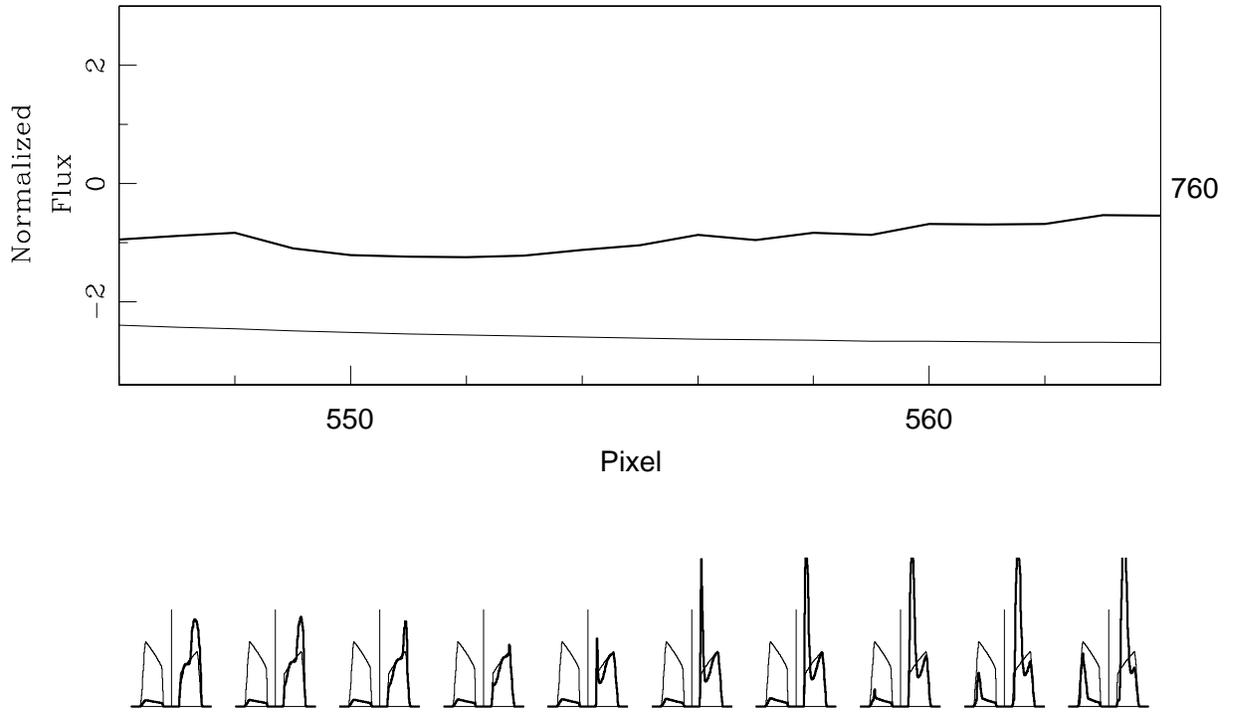}
\caption{The line profiles for the BLR and the light-curves for the BLR and the continuum for a track superimposed on the magnification pattern (Fig. \ref{fig17}) are shown. The parameter are $i=45^{\circ}$, $\theta_c=5^\circ$, $p=2$ , $r_{\rm cont}=0.096\eta_0$, $r_{\rm ex}=1 \eta_0$ and $r_{\rm in}=0.55\eta_0$. In the line profile panel the heavy solid lines represent the amplified line profile and the lighter solid lines correspond to the line profile normalized to mean amplification in the magnification map. In the light-curve panel the solid line represents the light curve from the BLR and the lighter solid line corresponds to the light-curve from the continuum. \label{fig20}}
\end{figure}

If the global displacement of the BLR is perpendicular to the shear, the time-scale of  the events depends exactly on the separation between stripes. In Fig. \ref{fig19} we have represented the BEL profiles with $r_{\rm ex}=1\eta_0$ corresponding to $20\times 20$ locations regularly distributed in the magnification pattern corresponding to a 10\% mass fraction in stars (see Fig. \ref{fig17}). It is clear that a rich phenomenology of line profile variability (including recurrent or alternating enhancement of parts of the line profile) can be reproduced in relatively short displacements across the magnification pattern. Inspectioning the magnification patterns it is possible to find paths to roughly reproduce two consecutive events of relative enhancement of one of the peaks without appreciable changes in the continuum. That considered in Fig. \ref{fig20}, for instance, implies a displacement across the pattern of 0.1 Einstein radii between the third and the sixth line profiles from left to right; that is, $\sim2$ years with $v_{\perp}=700$ kms$^{-1}$ (Oguri et al. 2004), according to the estimates by Richards et al. (2004b), but assuming 0.5$M_\odot$ stars in our case. Then the time between two consecutives line profiles is 0.62 years. The time between events of the order of 1 year can be obtained by considering that the mean mass of the stars is below 0.5$M_\odot$ or by supposing a factor two greater relative velocity between the quasar host and the lens galaxy, or by combining both possibilities.

\begin{figure}
\figurenum{21}
\plotone{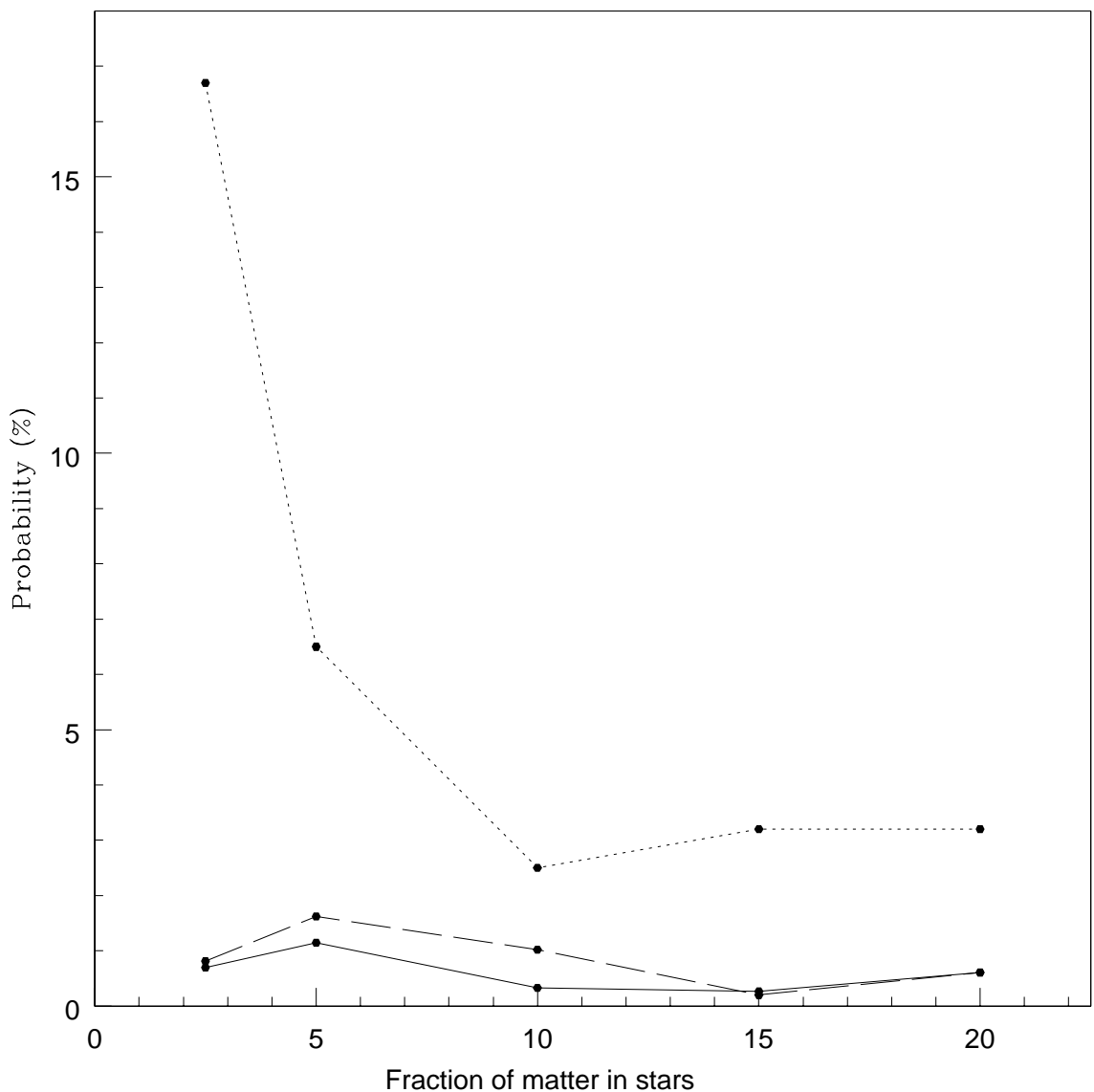}
\caption{The probability of founding repetitive line variations over short time-scales without considerable continuum's changes in our simulated magnification maps is represented versus the fraction of matter in stars. The dot line represents the probability of founding a track where the variability of the continuum light curves are below 0.15 magnitudes. The dash and solid lines represent the probability of founding a track where the variability of the continuum light curves are below 0.15 magnitudes, and repetitive line variability greater than 0.3 magnitudes, taking into account the peak and the flux intensity respectively (See text). \label{fig21}}
\end{figure}

To study the probability of founding repetitive line variations over short time-scales without considerable continuum changes we have computed and statistically studied light curves on the  magnification patterns corresponding to different fractions of matter in stars (2.5\%, 5\%, 10\%, 15\%, and 20\%) . We consider tracks of 20 pixels corresponding to 2 years (with $v_{\perp}=700$ kms$^{-1}$). We have extracted 1479 tracks from each magnification pattern. In first place, we look for continuum light curves with variability below 0.15 magnitudes. The probability of founding a track with this condition for the different patterns is represented in Figure \ref{fig21}. In second place, we identify, among these tracks, the cases in which at least one of the peaks follows a cycle: enhancement-fading-enhancement (if both peaks follow the cycle we can consider that one of them is strongly obscured) with an umbral of variability of at least 0.3 magnitudes (we have studied the variability of both the flux and the peak intensity). The probability of founding a track with both conditions, continuum variability below 0.15 magnitudes and repetitive line variability greater than 0.3 magnitudes, almost reach the 2\% for the magnification pattern generated with a 5\% fraction of matter in stars (see Fig.~\ref{fig21}). Thus, the probability of observing this event is small but not neglectable.

Figure \ref{fig19} also illustrates other very noticeable effects that a high density of caustics (originating from the high amplification and shear of J1004+4112) can induce in the line profiles generated by a bicone transversally oriented and transversally moving respect to the shear (and, hence, to the caustics). Because of the dependence of the velocity field with distance, each caustic will mainly affect emitters at a given velocity, enhancing a small part of the line profile. When more than one caustic  affects the bicone, several ``spikes'' can be seen in the line profiles (see Fig. \ref{fig19}). The displacement of the bicone across the caustics would result in the scanning of the kinematic properties of the BLR.

Thus, the  main trends of the spectral variability observed in J1004+4112 could be reproduced by the microlensing of a bicone moving transversally to the shear. The similarity of the computed phenomenon to the observed line variability can probably be improved by changing the proportion of matter in compact microlenses, by adjusting some of the model parameters (such as continuum and BLR sizes and separations) or by considering oblique paths. Independently of the details, the qualitative agreement between the proposed model and the observations in J1004+4112 supports the existence of an anisotropic gaseous system (susceptible of microlensing magnification) in the BLR that can give rise to substructure like shoulders or bumps in the line profiles. In the adopted model the inner and outer radii of the bicone are 9.1 and 16.6 l-days, respectively. These values are comparable to the value inferred from Kaspi et al. relationship for the high ionization lines in J1004+4112  (7.6 l-days).

Statistical studies make plausible the detection of an event with features like the ones observed in SDSS J1004+4112 but improbable the future detection of similar events. In the simulations, the probability is greater for the pattern generated with a 5\% fraction of matter in stars.

\section{CONCLUSIONS}

We have explored the effects of microlensing at high optical depth on a biconical model of arbitrary inclination and aperture for the BLR. We point out some results that could be of interest:

1 - For symmetric geometries the size of the BLR is directly related to its projected area. However,  a biconical BLR of any intrinsic size can have a small projected area if the cone aperture is small and the cone axis points toward the observer. This implies that very large amplifications, even greater than that of the continuum, are possible in microlensing events of a biconical BLR. For intermediate inclinations of the cone axis, $0^{\circ}<i<90^{\circ}$, microlensing induces relative enhancements between the red and blue parts of the line profile.

2 - Microlensing effects (global amplification, wavelength-dependent amplification), are related to the orientation of the projected bicone with respect to the shear of the magnification pattern (this could result in quasi-periodic effects when the source moves along the magnification pattern). In particular, the correlation between the continuum and BLR light-curves can be weak when the axis of the projected bicones is perpendicular to the shear.

3 - When the bicone is oriented transversally to a caustic, the emission line profile presents strong enhancement of a small wavelength region (spike). The movement of the bicone with respect to the caustic would scan the kinematics of the BLR. When the bicone lies in a caustic cluster, the emission line profile can be affected by several ``spikes''.

4 - We propose that the BLR can be composed by two gaseous systems. One more or less symmetrical that gives rise to the core of the line and other anisotropic,  the bicone, that contributes to the blue and red sides of the line. Under this hypothesis, the main features related to spectral variability detected in image A of SDSS
J1004+4112 (recurrent blueward enhancement, absence of continuum variability, and small time-scale) are qualitatively reproduced by microlensing of the biconical component moving perpendicular to the shear. The probability of an event reproducing these features is a $\sim2$\% in the most favorable case, corresponding to a fraction of mass in stars of 5\%. This probability is small but  makes plausible the detection of one event under the hypothesis of microlensing magnification of a bicone. However, future repetitions of the event are improbable under this hypothesis. Finally, it is noticeable that we can reproduce the observed variability with a bicone of inner and outer radii of 9.1 and 16.6 l-days, respectively. These values are comparable to the expected emissivity weighted radius (7.6 l-days) according to the statistical study of Kaspi et al. (2005).

\acknowledgments

We thanks the anonymous referee for valuable comments and suggestions. We acknowledge the Spanish Department of Education and Science grants AYA2004-08243-C03-01,
AYA2004-08243-C03-02 and
AYA2004-08243-C03-03 for financial support.
This work was also supported by the European Community's Sixth Framework Marie Curie Research Training Network Programme, Contract No. MRTN-CT-2004-505183 ``ANGLES''.

\appendix

\section{BLR models}
\label{apen}

To compute the integral in Eq.~\ref{fluxl}, we define $f$ as:

\begin{equation}
\label{fdef}
f \equiv \lambda - \lambda_0 \left [ 1 + \frac{v_0}{c} \left (
\frac{r}{r_{\rm in}} \right )^p \xi_{\perp,\parallel} \right ].
\end{equation}
and then, after integrating in the $r$ dimension and
adopting $x=c~(\lambda-\lambda_0)/( v_{\rm max}\lambda_0)$ 
and $x_{\rm m}=v_0/v_{\rm max}=(r_{\rm in}/r_{\rm ex})^p$, Eq.~\ref{fluxl}
becomes:
\begin{equation}
 F_x  = \left\{
  \begin{array}{ll}
        \int\limits_{0}^{2\pi}\left (\left [
        \int\limits_{0}^{\theta_{\rm c}} +
        \int\limits_{\pi-\theta_{\rm c}}^{\pi}\right ]
        \left [\epsilon(r)  r^2  \mu(\vec{r})
	\left ( \frac{\rm df}{dr}\right )^{-1}\right ]_{\rm f=0}  
	~\sin\theta d\theta \right )d\varphi,
  & \mbox {\small for $x_{\rm m}|\xi_{\perp,\parallel}|<|x|<|\xi_{\perp,\parallel}|$} \\
0 & \mbox {in other cases.}
    \end{array}
  \right.
\end{equation}

Moreover,
\begin{equation}
\label{renf}
[r]_{\perp,\parallel}^{\rm f=0} = r_{\rm in} \left ( \frac{x}{\xi_{\perp,\parallel} x_{\rm m}} 
\right )^{1/p},
\end{equation}
and
\begin{equation}
 \left [\frac{\rm df}{\rm dr}\right ]_{\rm f=0}  = \left |
     -  \frac{\lambda_0~p~v_0}{ r_{\rm in}~c} \left (
     \frac{[r]_{\perp,\parallel}^{\rm f=0}}{r_{\rm in}} \right )
     ^{p-1} \xi_{\perp,\parallel} ~\right |.
\end{equation}

\clearpage

\clearpage

\end{document}